\documentclass{aa}
\usepackage{graphicx}
\begin{document}
   \title{The 3-D ionization structure of the planetary nebula NGC~6565
\thanks{
Based on observations made with ESO Telescopes at the La Silla 
Observatories, under programme ID 65.I-0524, and on  
observations made with the NASA/ESA Hubble Space Telescope, 
obtained from the data archive at the Space Telescope Institute 
(observing program GO 7501; P.I. Arsen Hajian). 
STScI is operated by the association of Universities for Research in 
Astronomy, Inc. under the NASA contract  NAS 5-26555.  We have
applied the photoionization code CLOUDY, developed at the Institute of
Astronomy of the Cambridge University.}
}

\titlerunning{ The Planetary Nebula NGC~6565}

  \author{M. Turatto \inst{1} \and E. Cappellaro \inst{1} \and
  R. Ragazzoni \inst{1} \and S. Benetti \inst{1} \and F. Sabbadin
  \inst{1}}

   \offprints{M. Turatto, turatto@pd.astro.it}
   
   \institute{Osservatorio Astronomico di Padova, vicolo
   dell'Osservatorio 5, I-35122 Padova, Italy}
    
\authorrunning{Turatto et al.}

\date{}
  
\abstract{A detailed study of the planetary nebula
NGC~6565 has been carried out on long-slit echellograms 
($\lambda$/$\Delta\lambda$=60000, 
spectral range=$\lambda\lambda$3900--7750\AA) at six, equally spaced  position angles.  
The expansion velocity field, the
c(H$\beta$) distribution and the radial profile of the physical
conditions (electron temperature and density) are obtained. The
distance, radius, mass and filling factor of the nebula and the
temperature and luminosity of the central star are derived. The radial
ionization structure is analyzed using both the classical method and
the photo-ionization code CLOUDY.  Moreover, we present the spatial
structure in a series of images from different directions, allowing
the reader to ``see'' the nebula in 3-D.  
NGC~6565 results to be a young
(2000--2500 years), patchy, optically thick triaxial ellipsoid (a=10.1
arcsec, a/b=1.4, a/c=1.7) projected almost pole-on. The matter close to 
major axis was swept-up by some accelerating agent (fast wind? ionization? 
magnetic fields?), forming two faint and asymmetric
polar cups.  A large cocoon of almost neutral gas completely embeds
the ionized nebula. NGC~6565 is in a recombination phase, because of
the luminosity drop of the massive powering star, which is reaching
the white dwarf domain (logT$_*\simeq$5.08 K; logL$_*$/L$_\odot\simeq$2.0).  
The stellar decline started about 1000 years
ago, but the main nebula remained optically thin for other 600 years
before the recombination phase occurred.  In the near future the
ionization front will re-grow, since the dilution factor due to the
expansion will prevail on the slower and slower stellar decline.
NGC~6565 is at a distance of 2.0($\pm$0.5) Kpc and can be divided into
three radial zones: the ``fully ionized'' one, extending up to
0.029--0.035 pc at the equator (0.050 pc at the poles), the
``transition'' one, up to 0.048--0.054 pc (0.080 pc), the ``halo'',
detectable up to 0.110 pc.  The ionized mass ($\simeq$0.03 M$_\odot$)
is only a fraction of the total mass ($\ge$ 0.15 M$_\odot$), which has
been ejected by an equatorial enhanced superwind of
4($\pm$2)$\times$10$^{-5}$ M$_\odot$ yr$^{-1}$ lasted for
4($\pm$2)$\times$10$^3$ years.
}
\maketitle

%

\section{Introduction} 

\begin{figure*} \centering
\caption{WFPC2 appearance
of NGC~6565 in [OIII] (left) and [NII] (right). The images have been retrieved
from the HST archive. North is up and East is to the left.}  
\end{figure*}

The late evolution of low and intermediate mass stars 
(1.0 M$_\odot<$M$_*<$8.0 M$_\odot$) 
is characterized by the planetary nebula (PN) metamorphosis: an
asymptotic giant branch (AGB) star gently pushes out the surface layers in a
florilegium of forms, crosses the HR diagram and reaches the white dwarf regime 
(Aller 1984, Pottasch 1984, Osterbrock 1989).

So far the interpretation of the exuberant morphologies exhibited by the 
PNe in terms of detailed three-dimensional structures and physical 
conditions of the ionized gas was limited by projection effects, leading to
approximate spatial forms and to unrealistic assumptions for the main  
parameters, e.g. electron temperature and electron density constant 
all over the nebula (Aller 1984, 1990, 1994).

A 3-D reconstruction technique for studying at large and small scales the
morphology, physical conditions, ionization, spatial structure and evolutionary
status of PNe has been introduced by Sabbadin et al. (2000a, b) and Ragazzoni
et al. (2001), based on echellograms of moderate spectral resolution 
(R$\sim22000-25000$). The key of the 3-D methodology is simple: the PN is an
expanding plasma. Thus the position, thickness and density of each
elementary volume can be derived from the radial velocity, width
and flux of the corresponding emission.

The procedure has been applied to:

\begin{description}
\item[-] NGC~40: an optically thick, very low excitation barrel-shaped
nebula with thin arcs emerging at both ends of the major axis, powered by
a luminous and ``cold'' WC8 star presenting a large mass-loss rate. The
fast, hydrogen depleted photospheric material ejected by the nucleus is
gradually modifying the chemical composition of the innermost nebular
regions (Sabbadin et al. 2000a);
\item[-] NGC~1501: an evolved, high excitation, optically thin oblate
ellipsoid, denser in the equatorial belt, deformed by several bumps,
embedded in a homogeneous, inwards extended cocoon and ionized by a ``
hot'' and luminous WC4 star exhibiting nonradial g-mode pulsations 
(Sabbadin et al. 2000b; Ragazzoni et al. 2001).
\end{description}

In order to deepen the analysis, 
we have started a survey at high spectral and spatial resolutions with the ESO NTT. 
The superb quality 
of this material allows us to study at unprecedented accuracies 
objects with different morphology, e.g. NGC~7009 (the "Saturn" nebula),
the tetra-lobed IC~4634, the butterfly HB~5, Mz~3 and NGC~6537, 
the double-envelope NGC~5882, NGC~6153 and NGC~6818.

To simplify the application of the 3-D method we decided to begin
with an "easy" nebula without FLIERS (fast, low ionization emitting
regions), BRETS (bipolar, rotating, episodic jets), ansae, wings, multiple
envelopes etc., and naively selected NGC~6565.

\begin{figure}
   \caption{[NII]/[OIII] distribution over NGC~6565 (original frames
   and orientation as in Fig. 1), showing the large stratification
   effects and the faint, low ionization regions protruding from the
   main nebula in PA$\simeq$145$\degr$.}  
\end{figure}

\section{The nebula}

The bright, compact PN NGC~6565 (PNG~003.5-04.6, Acker et al. 1992) is ``a 
minute oval ring 10$\arcsec$x8$\arcsec$ in p.a. about 5$\degr$. Considerably fainter 
along the major axis, and the center is relatively vacant`` (Curtis 1918). 
Its HST appearance is shown in Fig. 1:
the [OIII] emission forms a quite homogeneous oval ring, sprinkled by a 
number of dark globules and spikes, mainly in the Northern sector. It is 
surrounded by a sharp, patchy [NII] layer presenting two faint 
lobes in PA$\simeq$145$\degr$.  Note also the light N-S asymmetry 
of the nebula with respect to the (barely visible) central star. The large stratification 
effects and the 
external, faint, low ionization regions are better seen in Fig. 2, presenting 
the [NII]/[OIII] distribution.

NGC~6565 appears peculiar in many respects:
\begin{description}
\item[-] despite the simple elliptical shape (Stanghellini et al. 1993; Gorny et 
al. 1997), it shows a rich low ionization spectrum typical of 
bipolar PNe (Greig 1972) and is N overabundant (Aller et al. 1988);
\item[-] the infrared  and radio excesses suggest the presence 
of a large amount of dusty neutral gas causing a detectable absorption 
of the optical radiation (Gathier et al. 1986; Stasinska et al. 1992);
\item[-] in spite of the high surface brightness of the nebula, indicative of an early evolutionary 
phase, the central star 
is faint and hot, i.e. it is in, or close to, the white dwarf domain.
Following Tylenda (1986), NGC~6565 could be in a recombination phase, 
due to the recent luminosity drop of the powering star.
\end{description}

Although the nebula is sometimes listed among the distance calibrators (Pottasch
1984; Sabbadin 1986; Gathier 1987; Zhang 1995), its distance is still poorly
defined and the distance-extinction law gives contradictory results (Gathier et al.
1986, Maciel et al. 1986 and de Oliveira-Abans \& Faundez-Abans 1991).

This paper is structured as follows: Sect. 3 presents the observational 
material and the reduction procedure, Sect. 4 is dedicated to the 
gas kinematics, Sect. 5 contains the H$\alpha$/H$\beta$ spectral maps, Sect. 6
analyses the radial distribution of the physical conditions (electron 
temperature and electron density), in Sect. 7  we derive the distance, the 
nebular mass and the central star parameters, in Sect. 8 the radial ionization 
structure and the distribution of the gas (ionized and neutral) are presented,
 Sect. 9 concerns the application of the photo-ionization model (CLOUDY),
Sect. 10 describes the 3-D structure of the nebula in different ions, a
short discussion and the conclusions are presented in Sect. 11.

\section{The observational material and reduction technique}

Six long-slit echellograms of NGC~6565 (exposure time 900s; spectral
resolution 60000 with a slit 1.0 arcsec wide) have been obtained on
July 28, 2000 at the ESO NTT equipped with EMMI in REMD mode using
grating \#14, grism \#3 and CCD \#36 (spectral range
$\lambda\lambda$3900-7750\AA, mean spatial scale along the slit 0.27
arcsec pix$^{-1}$).  During the observations the sky transparency was
excellent and the seeing fluctuated between 0.53 and 0.89 arcsec.  The
30 arcsec long slit was centered on the nebula for all the selected PA,
ranging from 0$\degr$ to 150$\degr$ with a constant step of 30$\degr$.

The reduction method follows conceptually the standard procedure,
including bias, flat field, distortion correction and wavelength and
flux calibration.

The slit length, much longer than the order separation,
leads to a considerable superposition of the echelle orders. This, on
the one hand does not affect the emission line spectrum of NGC~6565, on
the other hand it does the flat field exposures. Thus, we were forced
to apply a zero-order correction using a white-light flat
field. Thanks to the excellent CCD uniformity, the pixel to pixel
sensitivity fluctuation left by this procedure is below 5\% (verified
on the night sky and the comparison spectrum lines). All things
considered, it is a modest price to pay for keeping the whole spatial
information of the PN emissions in the $\lambda\lambda$3900-7750\AA\/
spectral range.

  \begin{figure*}
   \centering
      \caption{Main part of the NTT+EMMI frame of NGC~6565 at PA=150$\degr$ 
(in logarithmic scale); the principal nebular emissions are identified 
and the strongest 
night sky lines (N.S.) indicated. The spatial scale is given by the 
N.S. extension, corresponding to the slit length of 30$\arcsec$. }
   \end{figure*}

  \begin{figure*} \centering
  \caption{Detailed
  structure (in logarithmic scale) of some representative emissions
  observed in NGC~6565 at PA=150$\degr$. The orientation is as in Fig. 3.
  The fluxes are multiplied for the factor
  given in parenthesis to render each emission comparable with
  $\lambda$5007\AA\/ of [OIII]. The $\lambda$6300.304\AA\/ night sky
  line was removed from the [OI] nebular emission. Part of the [OI]
  $\lambda$6363\AA\/ emission (echelle order 96) appears as a
  moustache in the lower-left corner of the $\lambda$6300\AA\/ image
  (echelle order 97). The fast, blue-shifted wisp in the lower-left
  corner of the [NII] and [SII] frames is real.  A hot pixel column grazes the 
H$\alpha$ emission. The zero-velocity
  pixel column (zvpc) and the central star pixel line (cspl),
defined in Sect. 4, are shown in the lower--right panel.}
\end{figure*} 

After bias and flat field correction, the orders containing one or more
emissions of physical interest were extracted and treated as single, long-
slit spectra.  The distortion correction and the wavelength calibration
were performed by means of Th--Ar lamp exposures. During the frame
registration, the various orders were scaled to the slit height at 
H$\alpha$, chosen as reference line. Echellograms of the spectrophotometric
standard star HR5501 (sampled at several positions along each order) were
used to obtain the flux calibration.

We stress the substantial difference between our observational procedure
and the one generally adopted, which introduces an interference filter
to isolate a single order. On the contrary,  we cover the whole $\lambda\lambda$3900-7750\AA\/ 
spectral range with a single exposure.

For illustrative purposes, in Fig. 3 we present the main part of the
NTT+EMMI frame of NGC~6565 at PA=150$\degr$, which includes the external, 
low ionization regions, and in Fig. 4 the detailed structure (at the same
PA) of $\lambda$4686\AA\/ of HeII, $\lambda$5007\AA\/ of [OIII],
$\lambda$5876\AA\/ of HeI, $\lambda$6300\AA\/ of [OI],
$\lambda$6563\AA\/ of HI, $\lambda$6584\AA\/ of [NII],
$\lambda$6731\AA\/ of [SII] and $\lambda$7135\AA\/ of [ArIII]. They
suggest that the faint, low ionization regions visible in Fig. 2 are
the projection of extended cups protruding from the main nebula.  Note
also the large stratification effects, the [OI] rises in the outermost
parts and, in spite of the blurred H$\alpha$ appearance due to thermal
motions plus expansion velocity gradient plus fine structure, the similarity
of the HI and [OIII] distributions.  Following Williams (1973), these
features are indicative of an optically thick nebula powered by a high
temperature star.

A first, qualitative picture of the NGC~6565 structure can already
be inferred from Figs. 1 to 4: the object is an ellipsoid with extended polar
cups, projected almost pole-on. The possible optical appearance of the
nebula seen equatorial-on is illustrated in Fig. 5, showing the PN NGC~6886
in [OIII] and [NII]. NGC~6565 and NGC~6886 are almost twins in many
respects: nebular spectrum, physical conditions, ionization structure,
evolutionary phase, luminosity and temperature of the central
star. Likely, they derive from similar progenitors.

\begin{figure*}
\centering
\caption{HST imaging of the PN NGC6886 in [OIII] and [NII], showing the
possible appearance of NGC~6565 seen almost equatorial-on.}  
\end{figure*}

\section{The gas kinematics}

We introduce two important definitions relative to
each emission line, which will be extensively used in the paper (they
are shown in the lower-right panel of Fig. 4):
\begin{description}
\item[1]) the zero-velocity pixel column (zvpc, see Sabbadin et 
al. 2000a, 2000b,  and Ragazzoni et al. 2001), corresponding to the recession 
velocity of the whole nebula, collects the photons of the ionized gas 
expanding perpendicularly to the line of sight;
\item[2]) the central star pixel line (cspl), parallel to the dispersion, represents 
the nebular material 
projected at the apparent position of the star,
whose motion is purely radial.
\end{description}

In a certain sense the zvpc and the cspl are complementary: the 
first is independent on the expansion velocity and gives only 
spatial information, the second measures the velocity field and 
contains only kinematical information.
By combining the zvpc and the cspl we will derive the overall 
spatio-kinematical properties of the nebula.

The expansion velocity (Vexp) of the ionized gas in NGC~6565 was
derived from the analysis of the cspl in the different ions. Note that
the cspl is the same in all the frames, the spectra being radially arranged.
Thus, in order to achieve the faintest emissions (like
$\lambda$5198--5200\AA\/ of [NI], $\lambda$5517--5537\AA\/ of [ClIII] and
$\lambda$7005\AA\/ of [ArV]), the six echellograms were combined.

The results are contained in Table 1, where the ions are put in order
of increasing ionization potential (I.P.). Besides the peak separation
(2Vexp) of the main nebula (column 4), in column 5 we give Vexp of the
faint, fast, blue-shifted component emitted by the approaching,
southern polar cup, clearly visible in Figs. 3 and 4.

\begin{table*}
\caption{Expansion velocity in the cspl and peak separation in the zvpc for NGC~6565. }
\begin{tabular}{ccccccccccccc}
\hline
Ion & $\lambda$      & I.P.  &&2V$_{\rm exp}$(main)&V$_{\rm exp}$(cup)& &\multicolumn{6}{c}{peak separation($\arcsec$)}  \\
\cline{5-6}   \cline{8-13}
& (\AA)              & (eV)  && (km/s)  & (km/s)       & & PA=0$\degr$ & PA=30$\degr$ & PA=60$\degr$ & PA=90$\degr$ & 
PA=120$\degr$ & PA=150$\degr$  \\
\hline
$[$NI$]$   & 5198    &  0.0  &&    70.9:&   --         & &  8.7: &  8.4: & 8.0:  &  7.6: &  8.0:  &  8.3: \\
$[$OI$]$   & 6300    &  0.0  &&    67.7 &   55         & &  8.7  &  8.4  & 8.0   &  7.6  &  8.0   &  8.3  \\
$[$SII$]$  & 6731    & 10.4  &&    65.0 &   54         & &  8.5  &  7.8  & 7.5   &  7.3  &  7.5   &  7.9  \\
$[$OII$]$  & 7319    & 13.6  &&    59.4:&   --         & &  7.8: &  7.7: & 7.2   &  7.0  &  7.4   &  7.8: \\
HI         & 6563    & 13.6  &&    55.8 &   48:        & &  7.1  &  6.7  & 6.2   &  6.0  &  6.2   &  6.5  \\
$[$NII$]$  & 6584    & 14.5  &&    60.6 &   54         & &  8.2  &  7.6  & 7.4   &  7.1  &  7.4   &  7.8  \\
$[$SIII$]$ & 6312    & 23.4  &&    57.5 &   --         & &  7.6  &  6.9  & 6.7   &  6.3  &  6.5   &  6.7  \\
$[$ClIII$]$& 5537    & 23.8  &&    57.9:&   --         & &  7.0: &  6.9: & 6.5:  &  6.1: &  6.3:  &  6.7: \\
HeI        & 5876    & 24.6  &&    58.7 &   52:        & &  7.5  &  6.8  & 6.5   &  6.2  &  6.5   &  6.8  \\
$[$ArIII$]$& 7135    & 27.6  &&    57.2 &   50         & &  7.3  &  7.0  & 6.8   &  6.3  &  6.5   &  7.0  \\
$[$OIII$]$ & 5007    & 35.1  &&    56.7 &   49         & &  6.9  &  6.4  & 6.3   &  6.0  &  6.3   &  6.4  \\
$[$ArIV$]$ & 4740    & 40.7  &&    32.4 &   --         & &  3.9  &  3.6  & 3.5   &  3.5  &  3.6   &  3.9  \\
$[$NeIII$]$& 3967    & 41.0  &&    55.2 &   --         & &  7.2  &  6.6  & 6.5   &  6.3  &  6.6   &  6.6  \\
HeII       & 4686    & 54.4  &&    44.2:&   42:        & &  4.5  &  3.5: & 3.8:  &  3.7  &  4.0   &  4.1  \\
$[$ArV$]$  & 7005    & 59.8  &&    17.0:&   --         & &   --  &  3.4: & 3.2:  &  2.7: &  3.0:  &  2.7: \\
\hline
\end{tabular}
\end{table*}

As the cspl refers to the kinematical properties of the (almost) polar
gas in NGC~6565, so the zvpc contains the spatial properties of the
equatorial matter. The intensity peak separations, 2r$_{\rm zvpc}$,
in the different emissions are presented in columns 6 to 11 of
Table 1.

Typical errors for 2Vexp(main) and 2r$_{\rm zvpc}$ are 1.0 km
s$^{-1}$ and 0.15 arcsec for the strongest forbidden emissions ([OI],
[SII], [NII] and [OIII]) to 3.0 km s$^{-1}$ and 0.3 arcsec for the
faintest ones ([NI], [OII], [ClIII] and [ArV]). The corresponding
inaccuracies for the recombination lines are: 2.0 km s$^{-1}$ and 0.2
arcsec for HI, 1.5 km s$^{-1}$ and 0.2 arcsec for HeI and 3.0 km
s$^{-1}$ and 0.3 arcsec for HeII. The error in Vexp(cup)
(column 5) is 3.0 km s$^{-1}$ in all cases.

The correlation between cspl and zvpc, evident in Table 1, is
summarized in Fig. 6, showing both Vexp and
r$_{\rm zvpc}$ vs. I.P.. For reasons of homogeneity, the
peak separations at different PA are normalized to the [OI] value.

NGC~6565 follows the Wilson's law (``the high-excitation particles 
show smaller separations than the low excitation particles'', Wilson 1950)
with some caveats:
\begin{description}
\item[-] the ``high ionization'' zone (I.P.$>$50 eV) is poorly defined,
containing 
only two emissions: $\lambda$7005\AA\/ of [ArV] is exceedingly faint
and $\lambda$4686\AA\/ of HeII consists of thirteen fine structure
components whose spectral distribution presents a noticeable
blue-shifted tail lowering the measurement reliability (in fact,
2Vexp(HeII) from $\lambda$6560\AA\/ is 38$\pm$3 km s$^{-1}$);
\item[-] the ``medium ionization'' region (20 eV$<$I.P.$<$50 eV) is
characterized by a steep velocity gradient. Vexp and r$_{\rm zvpc}$ in [ArIV] are
considerably lower than in the other ions, as found by Wilson (1950)
for NGC~3242 and NGC~7662. On the contrary, [NeIII], a ``medium-
high'' ionization species, is emitted in the ``medium-low'' excitation region,
as observed in NGC~6818, NGC~6886 and NGC~7662 by Wilson (1950), and in
NGC~6720 by Reay \& Worswick (1977). These peculiarities will be analyzed in
Sect. 9 (photo-ionization model);
\item[-] the ``low ionization'' zone (I.P.$<$20 eV) is steep and well 
defined, with the obvious exception of hydrogen, which is emitted 
by the whole ionized nebula, independently on the plasma conditions. 
\end{description}

  \begin{figure*} \centering
  \includegraphics[angle=-90,width=\textwidth]{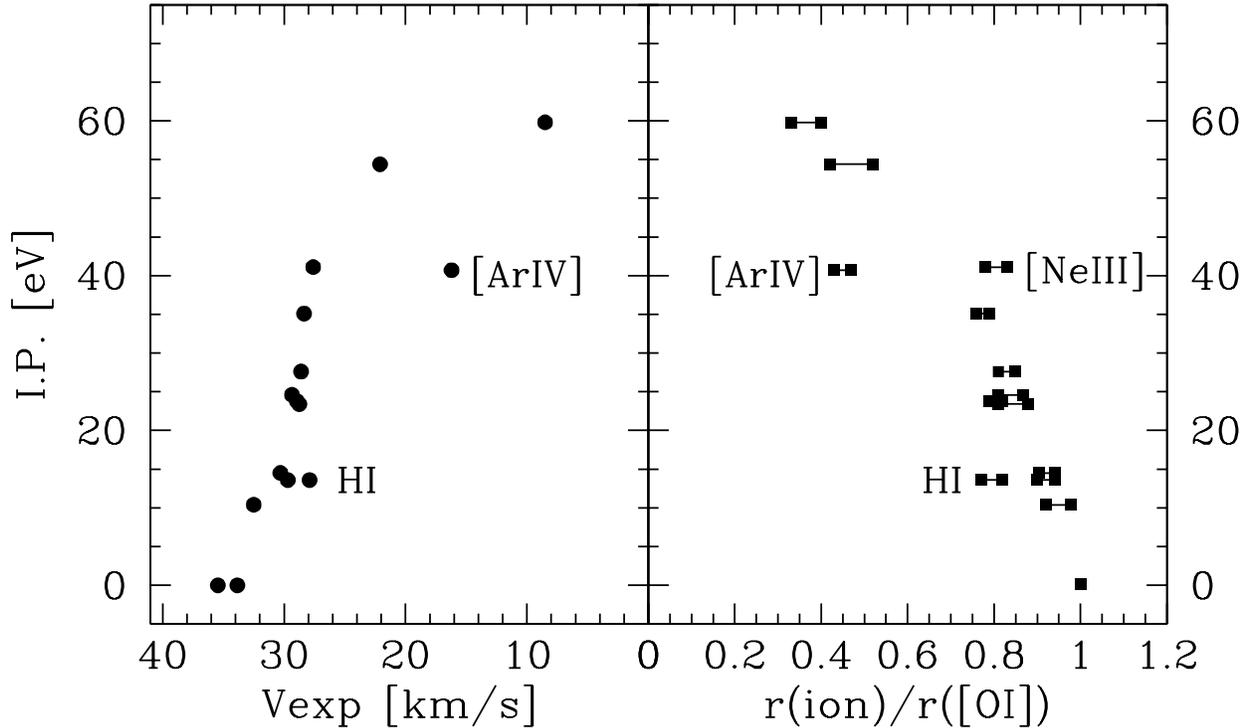}
  \caption{Polar expansion velocity (left) and equatorial peak
  separation (right) vs. I.P. in NGC~6565, showing the linear relation
  between Vexp and distance from the central star.  The intervals
  of the peak separations at different PA, normalized to the [OI] values,
  are reported. The ions 
  deviating from the normal (cfr. text) behavior are marked.}
  \label{FigVibSta} 
\end{figure*}

Assuming a minimum degree of symmetry within the
nebula we can assess that the expansion velocity is proportional to
the distance from the central star, in qualitative agreement with the
results of the classical papers of Wilson (1950) and Weedman (1968),
performed on a consistent sample of objects.
  
The exact value of the proportionality constant, A=Vexp/r$\arcsec$, remains
unknown, since at the moment no point of the nebula possesses both the 
expansion velocity and the distance from the star. Although a rigorous
treatment of the problem is presented in Sect.~\ref{neb_dist}, an
intuitive solution is evident in Fig. 4: by chance the combination of
spatial scale and spectral dispersion in the NTT+EMMI echellograms
gives emission line structures of NGC~6565 reproducing the tomographic
maps in the slice covered by the slit.

Let's come back to the gas kinematics. Further information is in the 
position--velocity maps, i.e. in the complete radial velocity field. 
The maps, shown in
Fig.~7 for the six PA, are in the nebula rest frame.  The systemic heliocentric
velocity is  Vr$_{\odot}=-5.2\pm0.5$ km s$^{-1}$. We have selected HeII, [OIII]
and [NII] as representative of high, medium and low ionization regions,
respectively, and H$\alpha$, which gives the overall distribution of the ionized
gas (Flux(H$\alpha$) $\propto$ Ne$^2$$\propto$N(H$^+$)$^2$, where Ne and
N(H$^+$) are the electron and proton densities, respectively).

  \begin{figure*}
   \centering
      \caption{Position-velocity maps for representative 
ions of NGC~6565 at the six PA observed. 
The orientation of each PA is indicated in the HeII panel. The contour step is $\Delta$logF=0.30. }
   \end{figure*}

A number of interesting features can be identified from these maps:
the complex expansion velocity field characterized by the
polar cups, the fast low excitation blue--shifted wisp in PA=0$\degr$ and
150$\degr$, the large stratification effects, the blurred  appearance
of H$\alpha$ due to a combination of thermal motions, fine-structure 
and expansion
gradient, and the blue-shifted tail in HeII (asymmetric distribution
of the thirteen fine-structure components).
In addition, our spectra point out the
presence of an external, very faint (untilted?) emission having
different kinematical properties with respect to the main nebula.
This ``halo'' (barely visible in Figs. 2, 3 and 7) is more evident in
[NII] than in H$\alpha$ and [OIII].  It appears double-peaked in [NII]
(2Vexp=25$\pm3$ km s$^{-1}$), with the blue-shifted component stronger
than the red-shifted one in the southern half of the nebula, and
vice-versa in the northern half. In both H$\alpha$ and [OIII], a
single, broad emission (FWHM=25$\pm5$ km s$^{-1}$) is visible.

The existence of a faint, low ionization envelope around 
NGC~6565 supports the Tylenda's (1983, 1986) suggestion that the 
nebula is in a recombination phase.

The only kinematical study of NGC~6565
reported in the literature is by Meatheringham et al. (1988), 
who derived 2Vexp=41.8 km s$^{-1}$ in [OII], 28.8 km s$^{-1}$ in [OIII] 
and 16.6 km s$^{-1}$ in HeII from the peak separation, 
and 2Vexp[OIII]=54.4 km s$^{-1}$
from the width at 10$\%$ maximum intensity (following Dopita et
al. 1985, this gives to a good approximation the highest expansion
velocity). 

The disagreement with the results of Table 1 can be ascribed 
to the different spatial and spectral accuracies 
(Meatheringham et al. 1988 used a 
slit width of 3.9$\arcsec$ and a spectral resolution $\simeq$26000), 
and to the different reduction techniques. 
To be noticed that the
width at 10$\%$ maximum intensity in our echellograms 
gives 2Vexp[OIII]=90($\pm$4) km s$^{-1}$. 
Further comments will be contained in subsection 10.2.

\section{The H$\alpha$/H$\beta$ spectral maps}

In general (see Aller 1984, Pottasch 1984, and Osterbrock 1989), the
PN extinction is obtained by comparing the
observed Balmer decrement (in particular H$\alpha$/H$\beta$) to the
intrinsic value given by Brocklehurst (1971) and Hummer \& Storey
(1987).  The estimates of (H$\alpha$/H$\beta$)$_{\rm obs}$ reported in the
literature for NGC~6565 span the range 3.79 (de Freitas Pacheco et
al.  1992) to 4.65 (Aller et al. 1988).  They all represent mean
values over the slice of nebula covered by the slit.

The spatial and spectral accuracies achieved by the
superposition technique used ($\pm0.15$ arcsec and $\pm1.0$ km s$^{-1}$,
respectively) allow us to extend the H$\alpha$/H$\beta$ analysis to
the whole spectral image.  Note that the fine structure of the lines
(the spectral distribution and the relative intensities of the
seven components of H$\alpha$ differ from the corresponding quantities
of the as many components of H$\beta$, Clegg et al. 1999) is
uninfluent for an expanding gas at Te$\simeq$10$^4$ K.

Fig. 8 shows the ${\rm H}\alpha/{\rm H}\beta$ isophotal contours
superimposed to the H$\alpha$ spectral emission (recall that
Flux(H$\alpha$)$\propto$N(H$^+$)$^2$) for some representative PA of
NGC~6565.
The large scale appearance of the Balmer ratio is the same in all the 
six spectra: there is a minimum (3.6--3.8) in the central regions,
it increases outwards (up to 4.4--4.6 just beyond the H$\alpha$ peaks),
then gradually decreases to 3.7--4.1 (the last statement is  
uncertain, because of the low signal--to--noise in the outer regions).   

So far, H$\alpha$/H$\beta$ variations were reported for a few PNe by
comparing direct, narrow-band imagery, and ascribed to the presence of
absorbing dust within (or around) the objects (for example, in NGC~6302
by Bohigas 1994, in NGC~6445 by Cuesta \& Phillips 1999, in NGC~2440 by
Cuesta \& Phillips 2000, and in NGC~6781 by Mavromatakis et al. 2001).

The advantage of a ``spectral map'' is evident: it shows the
de-projected slice of nebula covered by the slit, whereas the ``direct
images superposition'' refers to the PN projection on the sky. 

 \begin{figure*}
   \centering
      \caption{Isophotal contours of the H$\alpha$/H$\beta$ ratio
superimposed to the H$\alpha$ spectral image for some representative PA of 
NGC~6565. The orientation is as in Fig. 7. The isophotal contours range between 3.6 (the 
outermost) and 4.6, with a constant step of 0.2.} 
   \end{figure*}

In general, two main factors affect the H$\alpha$/H$\beta$ ratio:
\begin{description}
\item[a]) the amount of dust along the line of sight. 
We can write:
\begin{equation}
 c({\rm H}\beta)_{\rm tot}=c({\rm H}\beta)_{\rm interstellar} + 
c({\rm H}\beta)_{\rm local} 
\end{equation}
where c(H$\beta$)$_{\rm local}$ is null for optically thin, density bounded PNe. But
NGC~6565 is decidedly optically thick to the UV stellar radiation, mainly in the
dense, equatorial regions. Moreover, Gathier et al. (1986) and Stasinska et al.
(1992) suggest that the infrared and radio excesses of the nebula could be due
to a large amount of local absorbing material.
Our H$\alpha$/H$\beta$ spectral maps indicate that the neutral gas, if present, 
mainly surrounds the equatorial regions;
\item[b]) the Te distribution in the ionized gas: the higher the electron 
temperature, the smaller the intrinsic H$\alpha$/H$\beta$ ratio. For example, 
we have H$\alpha$/H$\beta$=2.74, 2.85 and 3.02 
for Te=20000 K, 10000 K and 5000 
K, respectively (assuming the case B of  Baker \& Menzel 1938, and 
log~Ne=3.00; Brocklehurst 1971, Aller 1984, Hummer \& Storey 1987).

A radial variation of Te within NGC~6565 is expected, due to the 
complexity of the ionization structure.
\end{description}

Note that a large amount of dust in the equatorial regions of the nebula should
produce spatial and wavelength-asymmetric H$\alpha$/H$\beta$ spectral maps,
whereas a radial Te variation should not. Unfortunately, our resolutions are
inadequate (by a factor of almost 2) to test such a possibility. In the present
case, the discrimination between a) and b) requires a detailed analysis of the
nebular physical conditions.

\section{The physical conditions}

The radial profiles of Te and Ne are obtained from the classical 
diagnostic line ratios (ions in p$^3$ configuration for Ne and in p$^2$ or 
p$^4$ configurations for Te; Aller 1984, Osterbrock 1989) and from the 
absolute H$\alpha$ flux distribution (Sabbadin et al. 2000a, b; Ragazzoni et 
al. 2001). 
A compilation of the relevant references for the transition probabilities 
and collision strengths used in this paper to derive Te, Ne and the 
ionic and chemical composition is given by Hyung \& Aller (1996) and Liu et al. (2000).

In all cases we make use of the zvpc, which is independent on the expansion 
velocity field, and corresponds to the equatorial, densest nebular regions of NGC~6565. 
The logical steps of our analysis are the following: 
\begin{description}
\item[-] first we obtain Te[OIII] from $\lambda$5007\AA/$\lambda$4363\AA\/ (the line ratio is almost 
independent on Ne for Ne$<$10$^4$ cm$^{-3}$);
\item[-]  Ne[SII] is then derived from  $\lambda$6717\AA/$\lambda$6731\AA\/ (using Te[OIII] to take into 
account the weak dependence of the ratio on Te); 
\item[-] Te[NII] comes from $\lambda$6584\AA/$\lambda$5755\AA\/ (adopting Ne[SII] for its weak dependence 
on Ne);
\item[-] last, Ne(H$\alpha$) is given by the H$\alpha$ flux distribution in the 
zvcp (both 
Te[OIII] and Te[NII] are considered for the Te dependence of the 
H$\alpha$ emissivity). 
\end{description}

The other diagnostics present in the spectra (like $\lambda$4711\AA/$\lambda$4740\AA\/ of 
[ArIV], $\lambda$5198\AA/$\lambda$5200\AA\/ of [NI], $\lambda$5517\AA/$\lambda$5537\AA\/ of 
[ClIII] and the [OII] red quartet around $\lambda$7325\AA) are too weak for an accurate spatial analysis.

This procedure, performed at all the six PA of NGC~6565, gives very similar results, as expected 
for an ellipsoid seen almost pole-on. Thus, 
only two representative PA are illustrated here: PA=0$\degr$ (along the 
apparent major axis) and PA=90$\degr$ (along the apparent minor axis).
The results are summarized in Fig. 9.

  \begin{figure*}
   \centering
   \includegraphics[angle=-90, width=\textwidth]{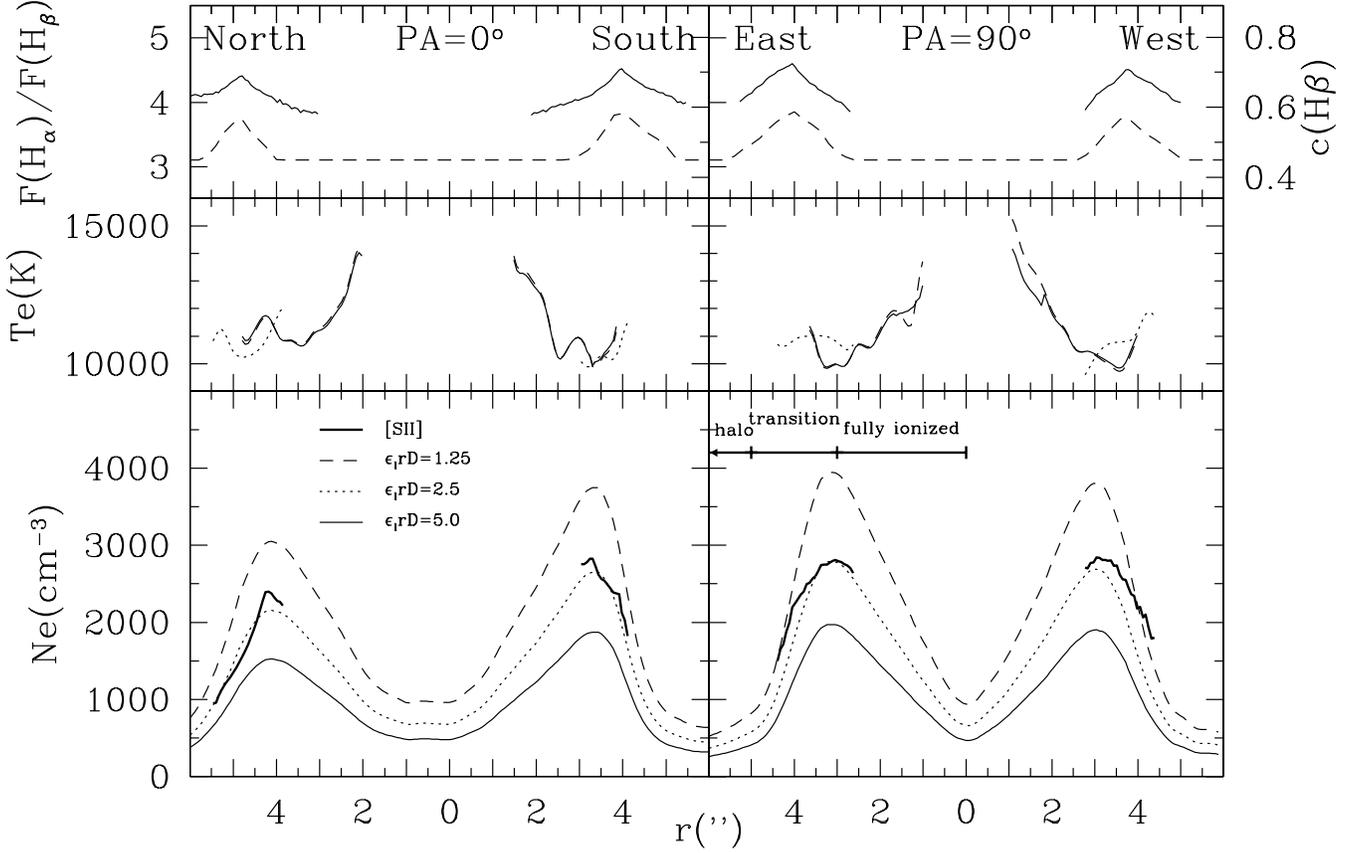}
\caption{Radial physical conditions and other parameters in NGC~6565 at PA=0
$\degr$ (left) and 90$\degr$ (right). Top panel: the observed H$\alpha$/H$\beta$
profile in the zvpc (left ordinate scale; continuous line) and the adopted 
c(H$\beta$)$_{\rm tot}$ profile (right ordinate scale; dashed line). Middle panel: Te[OIII] 
for cases A (i.e. c(H$\beta$) variable across the nebula; continuous line)
and B (c(H$\beta$)=constant=0.45; long--dashed line) and Te[NII] (dotted line).
Bottom panel: Ne[SII] (thick continuous line) and Ne from the H$\alpha$ flux for
some representative values of $\epsilon_{\rm l}\times$r$_{\rm cspl}$$\times$D (arcsec
Kpc). The three radial zones discussed in the text are indicated in the lower
right panel.}
   \end{figure*}

\subsection{Te[OIII]}

In order to get the intrinsic line ratio I($\lambda$5007\AA)/ I($\lambda$4363\AA), 
the observed line intensities must be
de-reddened. We have just found that H$\alpha$/H$\beta$ systematically changes
over NGC~6565 and, at the moment, we cannot identify the main cause for the
variation (Te or absorption).
Thus, we will consider two extreme cases:
\begin{description}
\item[A)] the H$\alpha$/H$\beta$ ratio only depends on the amount of absorption
along the line of sight. In practice this corresponds to assume a constant 
Te=10000 K;
\item[B)] the absorption is constant over the nebula and 
the H$\alpha$/H$\beta$ variation is due to Te. We discuss the case 
of c(H$\beta$)=0.45; the choice of other, reasonable 
values of c(H$\beta$) doesn't modify the conclusions here obtained.
\end{description}

The resulting Te[OIII] profiles, shown in Fig. 9, middle panel, appear
very similar in the two cases: the electron
temperature is large (14000--15000 K) in the innermost regions, it
rapidly decreases outwards down to about 10000 K, and then it slightly
increases (the last statement is weakened by the faintness of the [OIII]
auroral emission).  Previous Te[OIII] determinations in NGC~6565 are
mean values and range from 9700 K (Acker et al. 1991) to 10300 K
(Aller et al. 1988; de Freitas Pacheco et al. 1992).

The knowledge of the detailed Te[OIII] radial profile allows us to eliminate 
the dependence of the H$\alpha$/H$\beta$ ratio on Te, thus obtaining c(H$\beta$).
Our results indicate that:
\begin{description}
\item[-] c(H$\beta)$$_{\rm interstellar}$=0.45($\pm0.02$);
\item[-] in the innermost regions c(H$\beta$)$_{\rm local}\simeq$0. 
Once corrected for the interstellar component, the ratio
H$\alpha$/H$\beta$ is low, due to the large Te of the 
ionized gas. Te decreases outwards and, accordingly, H$\alpha$/H$\beta$ 
increases; 
\item[-] the H$\alpha$/H$\beta$ rise beyond the H$\alpha$ intensity peak is
mainly caused by the presence of neutral absorbing material around the ionized
nebula. The maximum of c(H$\beta$)$_{\rm local}$, 0.12($\pm0.03$), occurs at about
1.2($\pm$0.1)$\times$r$_{\rm H^+}$ (r$_{\rm H^+}$ being the radius of the H$\alpha$
peak).
\end{description}

The adopted, final c(H$\beta$)$_{\rm tot}$ profile for the two selected PA of NGC~6565 
is shown in Fig. 9, top panel. It will be used in the next Sections to 
correct the observed intensities through the relation: 

\begin{equation}
\log \frac{{\rm I}(\lambda)_{\rm corr}}{{\rm I}(\lambda)_{\rm obs}}=f_{\lambda}\, c({\rm H}\beta)
\end{equation}

where f$_{\lambda}$ is the interstellar extinction coefficient given by Seaton (1979). 

Notice that the adopted c(H$\beta$)$_{\rm tot}$ profile is obtained under the
assumption Te(H$^+$)$\simeq$Te[OIII]$\simeq$Te[NII]; this is questionable in the
outermost (knotty) regions, where the ionization drops and H$^+$ and H$^0$
co-exist.

Lowering Te(almost neutral gas), c(H$\beta$)$_{\rm local}$ decreases and goes to zero for 
Te(almost neutral gas)$\simeq$5000 K. In this case the overall 
H$\alpha$/H$\beta$ spectral distribution of Sect. 5 is due to Te variations. 
On the one hand this modifies only moderately the de-reddened line fluxes, 
on the other hand it implies that 
H$^+\gg$H$\degr$ (in other words: if the electron 
temperature of the almost neutral gas around NGC~6565 is $\simeq$5000 K, there is no neutral gas!). 
We will deep the question in the next Sections.

\subsection{Ne[SII]}

The intensity ratio $\lambda$6717\AA/$\lambda$6731\AA\/ of [SII] depends strongly 
on Ne and weakly on Te (Aller 1984; Osterbrock 1989). The accurate knowledge of
the c(H$\beta$) profile is unimportant, because the two lines are very close.
The resulting Ne[SII] distribution for the two selected PA of 
NGC~6565 is included in Fig. 9, bottom panel (thick continuous line). 
Our analysis suggests that:
\begin{description}
\item[-] the Ne[SII] profile is aligned to the H$\alpha$ one, and both peak
internally to the $\lambda$6717\AA\/ and $\lambda$6731\AA\/ maxima (which, 
in their turn, are lightly misaligned).
This indicates that the S$^+$ emission occurs in a recombining region;
\item[-] the Ne[SII] distribution shows peaks up to 2800$\pm$200 cm$^{-3}$; 
\item[-] the [SII] density peak is anticorrelated to r$_{\rm zvpc}$;
\item[-] the S$^+$ emitting zone, being co-spatial to the N$^+$ 
one (see Fig. 1), has a patchy and inhomogeneous structure favoring the 
detection of the densest, i.e. brightest, parts and indicating that the local filling factor, 
$\epsilon_{\rm l}$, in the external, low ionization layers of NGC~6565 is $\epsilon_{\rm l}$$<$1.
\end{description}

 Previous Ne[SII] determinations in the nebula (mean values) date from 
Aller et al. (1988, Ne=1600 cm$^{-3}$), Acker et al. (1991, Ne=1460 
cm$^{-3}$) and de Freitas Pacheco et al. (1992, Ne=1000 cm$^{-3}$). Moreover, Meatheringham et 
al. (1988) obtain Ne=1550 cm$^{-3}$ from $\lambda$3726\AA/$\lambda$3729\AA\/ of [OII].



\subsection{Te[NII]}

The line intensity ratio $\lambda$6584\AA/$\lambda$5755\AA\/ of [NII] depends
strongly on Te and weakly on Ne (Aller 1984; Osterbrock 1989). Adopting the 
Ne[SII] distribution we have derived the Te[NII] one shown in Fig.~9 (middle panel,
dotted line). Te[OIII] and Te[NII] are only partially superimposed: [OIII] is
present all over NGC~6565, whereas [NII] is peaked in the outermost, low
ionization regions. Their values are similar;  the electron temperature 
discrepancy, normally observed in PNe, can be ascribed to inaccuracy in the 
atomic parameters, small scale ionization fluctuation and, mainly, weakness of the 
auroral lines (Aller 1990; Gruenwald \& Viegas 1995; Mathis et al. 
1998).  
Previous Te[NII]
determinations in NGC~6565 are mean values and span in the range 8900--9500 K 
(Aller et al. 1988; Acker et al. 1991; de Freitas Pacheco et al. 1992).

\subsection{Ne from the H$\alpha$ flux}

Following Sabbadin et al. (2000a, b) and Ragazzoni et al. (2001), the accurate
Ne radial distribution can be obtained from the absolute H$\alpha$ flux in the
zvpc. The de-reddened H$\alpha$ profile must be further corrected for the
instrumental resolution,
thermal motions and fine structure (for details, see Sabbadin et al.
2000a). These corrections  were superfluous for the diagnostics just analyzed,
i.e. $\lambda$5007\AA/$\lambda$4363\AA, $\lambda$6717\AA/$\lambda$6731\AA\/ and
$\lambda$6584\AA/$\lambda$5755\AA, because they are forbidden lines and each
ratio refers to the same ion and is emitted in the same nebular region.

For a pixel of the H$\alpha$ zvpc we have:

\begin{equation}
4\pi D^2 F(H\alpha) = 4\pi j_{\rm H\alpha} N(H^+) Ne V_{\rm l} \epsilon_{\rm l}
\end{equation}
where:
\begin{description}
\item[-] D is the nebular distance;
\item[-] j$_{\rm H\alpha}$ is the emission coefficient (case B of Baker \& Menzel 
1938), interpolated from Brocklehurst (1971) values;
\item[-] V$_{\rm l}$ is the ``local volume'', i.e. the nebular volume identified by a single 
pixel of the zvpc; V$_{\rm l}$ is given by l$\times$b$\times$s$\times$(D/206265)$^3$, 
where l=slit width (1.0$\arcsec$), 
b=pixel spatial scale (0.27$\arcsec$) and s is the radial depth of the zvpc, 
s=r$_{\rm cspl}\Delta$V/Vexp, with $\Delta$V=pixel spectral resolution (1.79 km 
s$^{-1}$), Vexp=28 km s$^{-1}$ (see Table 1) and r$_{\rm cspl}$ is the (still 
unknown) angular radius of the cspl (i.e. the H$\alpha$ nebular size in the 
radial direction);
\item[-] $\epsilon_{\rm l}$, the ``local filling factor'', corresponds to the fraction of the local 
volume V$_{\rm l}$ which is actually filled by matter with density Ne. 
\end{description}

We can make the reasonable assumptions that Ne$=1.15\times$N(H$^+$) and 
$\epsilon$$_{\rm l}$=constant along the zvpc, thus obtaining:


\begin{equation}
Ne = \frac{1.07\times 10^9}{Te^{-0.47}} \times 
(\frac{F(H\alpha)}{\epsilon_{\rm l} \times r_{\rm cspl} \times D})^{1/2}
\end{equation}

Eq. (4), combined with Ne$\simeq$Ne[SII], represents an important link  between
the spatial and the dynamical properties of the ionized gas (i.e. between the
zvpc and the cspl). It can be used to derive the distance if independent
information on the radial radius, r$_{\rm cspl}$, is available (for example, 
$\epsilon_{\rm l}$=1 and r$_{\rm cspl}$=r$_{\rm zvpc}$ in a homogeneous PN presenting
a large degree of symmetry). In the case of NGC~6565 we have the opposite
situation: the distance will be obtained in the next Section by means
of the extinction-distance correlation; thus, Eq. (4) will be used to derive
$\epsilon_{\rm l}$ and the radial size of the cspl.

Figure 9, bottom panel, shows the Ne distribution for three representative
values of $\epsilon_{\rm l}\times$r$_{\rm cspl}$$\times$D (1.25, 2.5 and 5.0 arcsec Kpc).
Note the Ne radial asymmetry (the decline outward the maximum is steeper
than the internal rise), and the similarity with the [SII] density profile, the
alignment of the peaks and the close overlap for $\epsilon_{\rm l}\times$r$_{\rm cspl}$
$\times$D=2.5 arcsec Kpc.

\subsection{Physical conditions: general considerations}

Our analysis shows that the equatorial regions of NGC~6565 consist of three, 
distinct radial zones, indicated in the right-bottom panel of Fig.~9: 
\begin{description}
\item[-] the ``fully ionized'' nebula is the innermost one (roughly 
extending up to the H$\alpha$ peak). Ne 
monotonically increases outwards, while Te first decreases then 
remains almost constant;
\item[-] the ``transition'' zone located just beyond the electron density peak.
Here Ne rapidly decreases to a few hundreds cm$^{-3}$ and Te softly increases
outwards;
\item[-] the outermost nebula, the ``halo'', presenting a gentle gradient of Ne;
nothing can be said on Te$_{\rm halo}$.
\end{description}

Note that:
\begin{description}
\item[-] the overall Ne profile can be obtained from the H$\alpha$ flux in the
zvpc, whereas Te is limited to the main nebula, due to the weakness of the
[OIII] and [NII] auroral lines. Much deeper echellograms are necessary (and
highly advisable) to determine Te up to the faintest regions;
\item[-] the  Ne of the halo shown in Fig.~9 (bottom panel) must be considered as 
an upper limit since we have adopted the same kinematical properties for the
main nebula and the halo. A better approximation is given by:

\begin{equation}
Ne_{\rm halo} = \frac{Ne_{\rm halo}({\rm Fig. 9})}{ [\frac{r_{\rm halo}}{
r_{\rm main}}\times \frac{Vexp_{\rm main}}{Vexp_{\rm halo}}]^{1/2}} \simeq 
\frac{Ne_{\rm halo}({\rm Fig. 9})}{2}   
\end{equation}

\item[-] Te refers to the ionized gas, but
there are growing evidences indicating the presence of a considerable amount of
almost neutral, patchy gas (at unknown Te) in both the ``transition'' zone and
the ``halo'';
\end{description}

Up to now we have obtained the physical conditions in the equatorial regions of
NGC~6565 using the zvpc. The linear relation between expansion velocity and
nebular radius, found in Sect. 4, suggests that the same analysis can be
performed also in the polar regions. The main problem is represented by the line
fluxes, which are considerably lower in the cspl than in the zvpc.

Only a rough Ne profile can be derived for the (almost) polar regions, showing
density peaks of 1000($\pm200$) cm$^{-3}$ in [SII] and 500($\pm100$) cm$^{-3}$
from the H$\alpha$ flux (adopting Te=10$^4$ K and $\epsilon_{\rm l}$=1; see subsect.
7.2).

\section{The nebular distance, radius and mass and the central star parameters}
\label{neb_dist}

\subsection{Distance}

In Sect. 5 we have obtained c(H$\beta$)$_{\rm interstellar}$=0.45$\pm$0.02,
corresponding to E(B-V)=0.31$\pm$0.01 (Acker 1978). The introduction of this
value in the reddening-distance diagrams presented by Gathier et al. (1986,
weight 2), Maciel et al. (1986, weight 1) and de Oliveira-Abans \& Faundez-Abans
(1991, weight 1) gives a nebular distance of $2.0\pm0.5$ Kpc. Previous
individual estimates, based on lower values of c(H$\beta$), span the range 
0.5 to 1.55 Kpc,
whereas statistical distances are spread between 0.9 
(Gathier 1987) and 7.7 Kpc (Kingsburgh \& Barlow 1992).

Using the scheme introduced in the previous Section and D=2.0 Kpc we have:

R$_{\rm fully~ionized} = 0.035 \times 0.029$ pc 

R$_{\rm transition} = 0.054 \times 0.048$ pc  

R$_{\rm halo} \simeq$ 0.11 pc (indicative value).  

\subsection{Filling factor and r$_{\rm cspl}$}

Recalling, from sub-section 6.4, that the best match of Ne[SII] with 
Ne(H$\alpha$) occurs for $\epsilon_{\rm l}\times$r$_{\rm cspl}\times$D=2.5
arcsec Kpc, we immediately have $\epsilon_{\rm l}\times$r$_{\rm cspl}$=1.25 arcsec. Note
that for $\epsilon_{\rm l}\sim1$ we obtain ellipsoids compressed along the
polar axis (r$_{\rm cspl}$=1.25 arcsec, i.e. r(equatorial)/r(polar)$\simeq$2.6). 
Equally improbable are the solutions for 1$\gg\epsilon_{\rm l}$; 
for $\epsilon_{\rm l}$=0.1 we
have r$_{\rm cspl}$=12.5 arcsec, that is a nebula very stretched along the polar axis; 
r(equatorial)/r(polar) $\simeq$0.26.

A more stringent upper limit to $\epsilon_{\rm l}$, $\epsilon_{\rm l}$$\le$0.35, results
from the intuitive assumption r(polar)$\ge$r(equatorial), i.e. r$_{\rm cspl}\ge$r$_{\rm zvpc}$.
Moreover, by extrapolating to the polar regions the anticorrelation 
between Ne[SII](peak) and r$_{\rm zvpc}$ found for the equatorial zone 
(cfr. Sect. 6.2), 
we derive r$_{\rm cspl}$$\simeq$5.0 arcsec (and $\epsilon_{\rm l}$$\simeq$0.25). 
This is confirmed by the morphological analysis of the ``NGC~6565-like''
PNe projected more or less equatorial-on, contained in the catalogues of
Acker et al. (1992), Schwarz et al. (1992), Manchado et al. (1996), and Gorny et
al. (1999), e.g. NGC~6886 (PNG 060.1-07.7; see Fig. 5), M~2-40 (PNG 024.1+
03.8), K~3-92 (PNG 130.4+03.1), M~2-53 (PNG 104.4-01.6) and M~1-7 (PNG 189.8+07.7).

Thus, in the following we will adopt r$_{\rm cspl}$=5.0$\pm0.5$ arcsec 
(corresponding to 0.048$\pm$0.005 pc), and $\epsilon_{\rm l}$=0.25, 
constant across the S$^+$ emitting region (consistent with the patchy structure
of the low ionization layers). The position-velocity proportionality constant,
A=Vexp/r$_{\rm cspl}$, results to be 5.6$\pm$0.6 km s$^{-1}$ arcsec$^{-1}$; this
agrees with the impression that the NTT+EMMI emission
line structure of NGC~6565 nicely reproduces (to within 20$\%$) the tomographic
map of the nebular slice covered by the slit; in fact, in Fig. 3, 4, 7 and 8 one
arcsec corresponds to 3.11 pixels along the dispersion (cspl) and to 3.70 pixels
along the slit (zvpc).

The kinematical age of the nebula, t$_{\rm kin}$$\propto$A$^{-1}$$\simeq$1750 yr, represents a
lower limit to the actual age, $t_{\rm N6565}$, since the dynamical history of the gas
is unknown (Aller 1984; Schmidt-Voigt \& K\"oppen 1987, and Dopita et al. 1996).
An estimate of $t_{\rm N6565}$ can be obtained by assuming a nebular ejection at 
Vexp(superwind)$<$Vexp(N6565), followed by a constant acceleration up to 
Vexp(N6565). Vexp(superwind) can be measured in the OH/IR sources 
(i.e. stars surrounded by thick opaque envelopes of gas and dust producing 
maser emissions), which are
believed to be the PNe progenitors (see Habing 1996 and references therein). For
Vexp(superwind)=15($\pm$5) km s$^{-1}$ (David et al. 1993; Chengalur et al.
1993, and Sjouwerman et al. 1998) we derive t$_{\rm N6565}$=2300($\pm$300)yr. 

\subsection{Nebular mass}

Given the distance, the ionized nebular mass is obtainable in different ways 
(from the H$\beta$ flux, the radio flux and the observed Ne distribution; Aller
1984; Pottasch 1984; Osterbrock 1989). Adopting D$=2.0$ Kpc, we get M$_{\rm ion}$=0.03
($\pm0.01$) M$_\odot$, that is considerably lower than the mean nebular mass of
optically thin PNe, M$_{\rm PNe}$$\simeq$0.10--0.15 M$_\odot$ (Pottasch 1983; Boffi
\& Stanghellini 1994; Dopita et al. 1996).
This confirms that NGC~6565 is ionization bounded and suggests the presence of 
a considerable amount of surrounding, neutral gas. From simple 
geometrical considerations we can put  M$_{\rm PNe}$/M$_{\rm ion}$$\simeq$(R$_{\rm tot}$/R$_{\rm ion}$)$^4$, 
where R$_{\rm tot}$ is the radius of the (ionized 
+ neutral) density peak. We have R$_{\rm tot}$$\simeq$1.4$\times$R$_{\rm ion}$, 
i.e. similar to the radius of the peak of c(H$\beta$) 
given by the H$\alpha$/H$\beta$ spectral maps. 

An estimate of the neutral gas density, N(H$^0$), at R$_{\rm tot}$ can be
obtained from  E(B-V)$_{\rm local}$= c(H$\beta$)$_{\rm local}/1.48$ (Acker 1978), taking
a ``normal''  gas to dust ratio (Spitzer 1978) and assuming a radial thickness of the
absorbing layer comparable to R$_{\rm tot}$. The resulting value,  
N(H$^0)\simeq3(\pm1)\times 10^3$ cm$^{-3}$, agrees with all the previous
indications that a dense, neutral envelope surrounds the ionized nebula.

A further, independent confirmation comes from the detailed analysis of both 
the consistency and distribution of the dark globules superimposed on the 
[OIII] image of NGC~6565 (Fig.~1):
\begin{description}
\item[-] a number of intensity scans secured in the neighbourhood of the 
globules indicates a depletion of the [OIII] flux up to 30\%, corresponding 
to c(H$\beta$)$_{\rm globules}$ up to 0.11 (to be compared with 
c(H$\beta$)$_{\rm local}$=0.12$\pm$0.03 obtained in Sect. 5 from the 
H$\alpha$/H$\beta$ spectral maps);
\item[-] most globules appear in the northern part of the nebula, 
as expected of an ellipsoid denser in the equatorial regions and seen 
almost pole-on. Figs. 3, 4 and 7 show that the southern polar cup of NGC~6565 is approaching, 
thus the shadowing by the neutral, dense gas located in the equatorial 
regions mainly affects the northern portion of the image.
\end{description}

All this, added to the different general morphology of the nebula in [OIII] and 
[NII] (quite homogeneous and diffuse in [OIII], inhomogeneous and patchy in 
[NII]; Fig. 1) indicates the presence of noticeable density fluctuations at 
small spatial scale (below the 
seeing, $\simeq$0.7 arcsec) in the outermost, recombining nebular regions,
which are rich of knots, globules and condensations.

\begin{figure*}
\centering
\includegraphics[angle=-90,width=\textwidth]{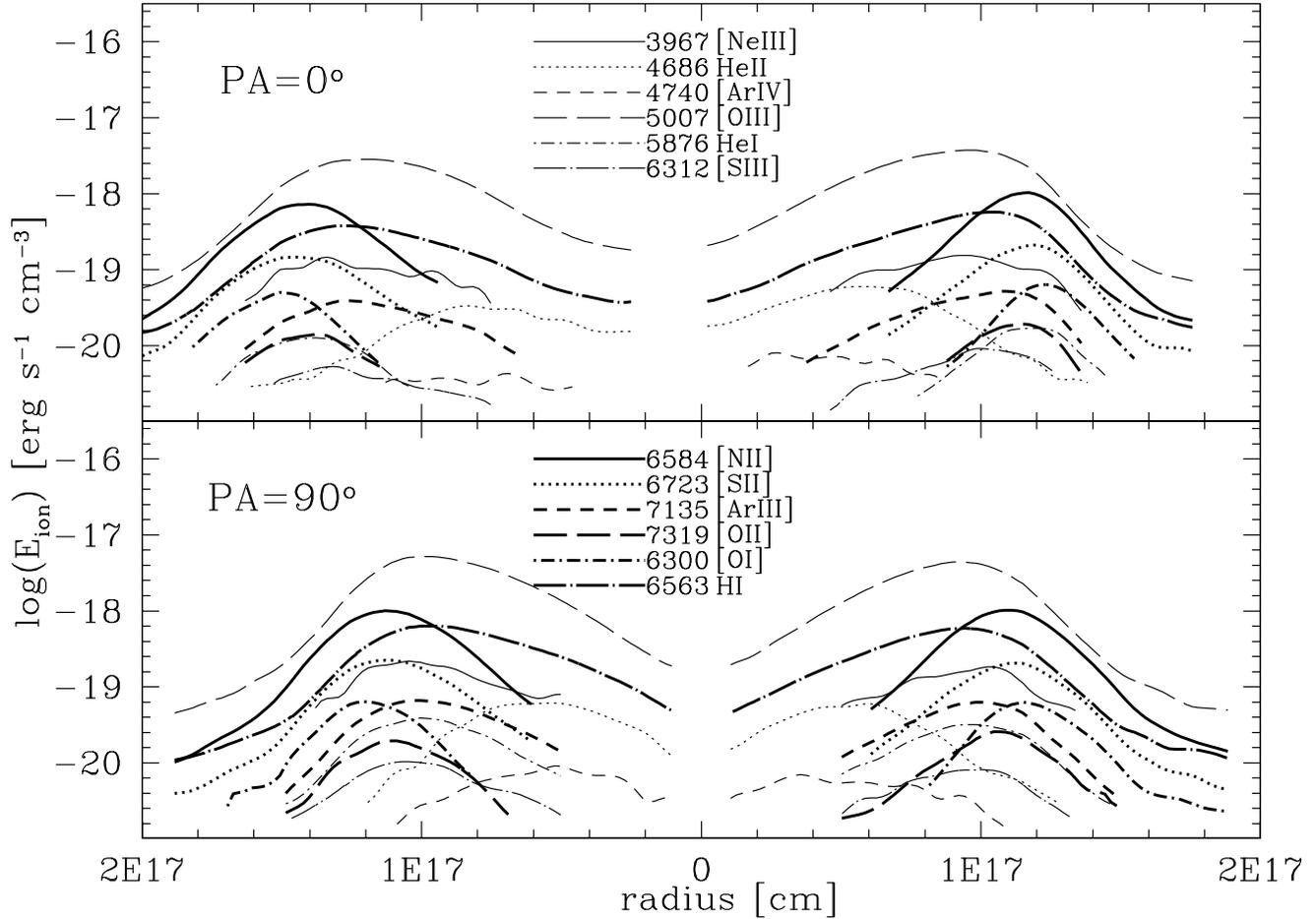}
\caption{Absolute emission line radial profiles (erg cm$^{-3}$ s$^{-1}$; 
logarithmic scale) in the zvpc of NGC~6565 at PA=0$\degr$ and 90$\degr$. 
The orientation is as in Fig. 9. The nebula has been considered  
at a distance D=2.0 Kpc.}
\end{figure*}

\subsection{Central star parameters}

The accurate value of m$_{\rm V*}$ has been obtained from 
WFPC2 frames of NGC~6565 taken through the broad-band filter F555W
($\lambda_{\rm c}$=5407\AA, bandwidth=1236 \AA). The resulting magnitude is
m$_{\rm V*}$=18.88($\pm0.05$), where the unknown star color is the main
source of inaccuracy. Previous m$_{\rm V*}$ estimates reported in the
literature span the range 15.9 (Shaw \& Kaler 1989) to 20.3 (Gathier
\& Pottasch 1988).

The HI and HeII Zanstra temperatures, T$_{\rm Z}$HI and T$_{\rm Z}$HeII
respectively, are derived in the classical way (Aller 1984; Pottasch
1984; Osterbrock 1989) by comparing the de-reddened H$\beta$ and HeII
$\lambda$4686\AA\/ nebular fluxes with (m$_{\rm V*})_{\rm o}$.  The integrated
H$\beta$ and $\lambda$4686\AA\/ fluxes were extrapolated from the
overall line profile at each PA, assuming a circular symmetry of the
nebular image.  We obtain log F(H$\beta$)$_{\rm obs}=-11.20(\pm0.03$)
mW$\times$m$^{-2}$ and
I($\lambda$4686\AA/H$\beta$)$_{\rm obs}$=0.12($\pm0.02$).
The corresponding values reported in the literature are in the range
-11.27 (photoelectric photometry; O'Dell 1962) to -11.1 (slit
spectroscopy; Acker at al. 1991) for log F(H$\beta$)$_{\rm obs}$, and 0.12
(photoelectric narrow-band photometry; Kohoutek \& Martin 1981) to
0.22 (slit spectroscopy; Acker et al. 1991) for
I($\lambda$4686\AA/H$\beta$)$_{\rm obs}$.

We derive log(T$_{\rm Z}$HI)=5.20($\pm0.05$) and log(T$_{\rm Z}$HeII)=
5.08($\pm0.05$), thus confirming the peculiarity already reported by
Pottasch (1981), Martin (1981), Reay et al. (1984), Gathier \&
Pottasch (1988, 1989) and Jacoby \& Kaler (1989): the Zanstra
discrepancy is reversed in the central star of NGC~6565, being
T$_{\rm Z}$HI$>$T$_{\rm Z}$HeII.

The stellar luminosities are logL/L$_\odot$(T$_{\rm Z}$HI)=2.3 ($\pm0.3$) 
and log L/L$_\odot$(T$_{\rm Z}$HeII)=2.0($\pm0.3$) (using the bolometric 
corrections by Sch\"onberner 1981).

The low luminosity and high temperature of the star, combined with the short 
age of the nebula, suggest that the stellar mass, M$_*$, is larger than 
the average value ($\simeq$0.60 M$_\odot$) of the PNe nuclei (Sch\"onberner 1979, 1981, 1983; 
Vassiliadis \& Wood 1993, 1994; Bl\"ocker 1995).
The quantification of M$_*$ through the comparison with
theoretical evolutionary tracks
reported in the literature gives no solutions for the He-burning 
post-AGB
stars (these models do not predict a fast luminosity decline), whereas for the
H-burning post-AGB stars different values are obtained:

\begin{description}
\item[]M$_*$$>$0.70 M$_\odot$ (Sch\"onberner 1981, 1983, Iben 1984), 
\item[]M$_*$$>$0.90 M$_\odot$ (Wood \& Faulkner 1986, Vassiliadis \& Wood 1994),
\item[]M$_*$$\simeq$0.65 M$_\odot$ (Bl\"ocker \& Sch\"onberner 1990, 
Bl\"ocker 1995),
\end{description}

These discrepancies reflect the different choices by the authors of:
\begin{description}
\item[-] the semi-empirical AGB mass-loss law, controlling both the 
evolution along the upper AGB and the transition into the PNe region, 
\item[-] the residual, hydrogen rich envelope (and its correlation with 
the thermal-pulse cycle phase), and the fast post-AGB mass-loss, 
determining the horizontal crossing of the HR diagram,
\item[-] the treatment of the gravo-thermal energy release and 
neutrino energy losses, defining the fading to the white dwarf regime.
\end{description}

Although a deep analysis of the central star evolution is beyond the aim of
this paper, our observational data suggest the following, qualitative picture:
the massive progenitor of NGC~6565 underwent a long AGB phase (favoring the
degeneration of the core), a strong superwind leaving a small, hydrogen rich
envelope (causing a rapid horizontal crossing), and a quite fast final fading
towards the white dwarf domain (highly degenerated core and neutrino
losses).

According to the critical analysis by Bl\"ocker \& Sch\"onberner (1990),
Tylenda \& Stasinska (1994) and Bl\"ocker (1995), we assume
M$_*$(NGC~6565)$\simeq$0.65 M$_\odot$.

\begin{figure*}
\centering
\includegraphics[width=13.7cm,angle=-90]{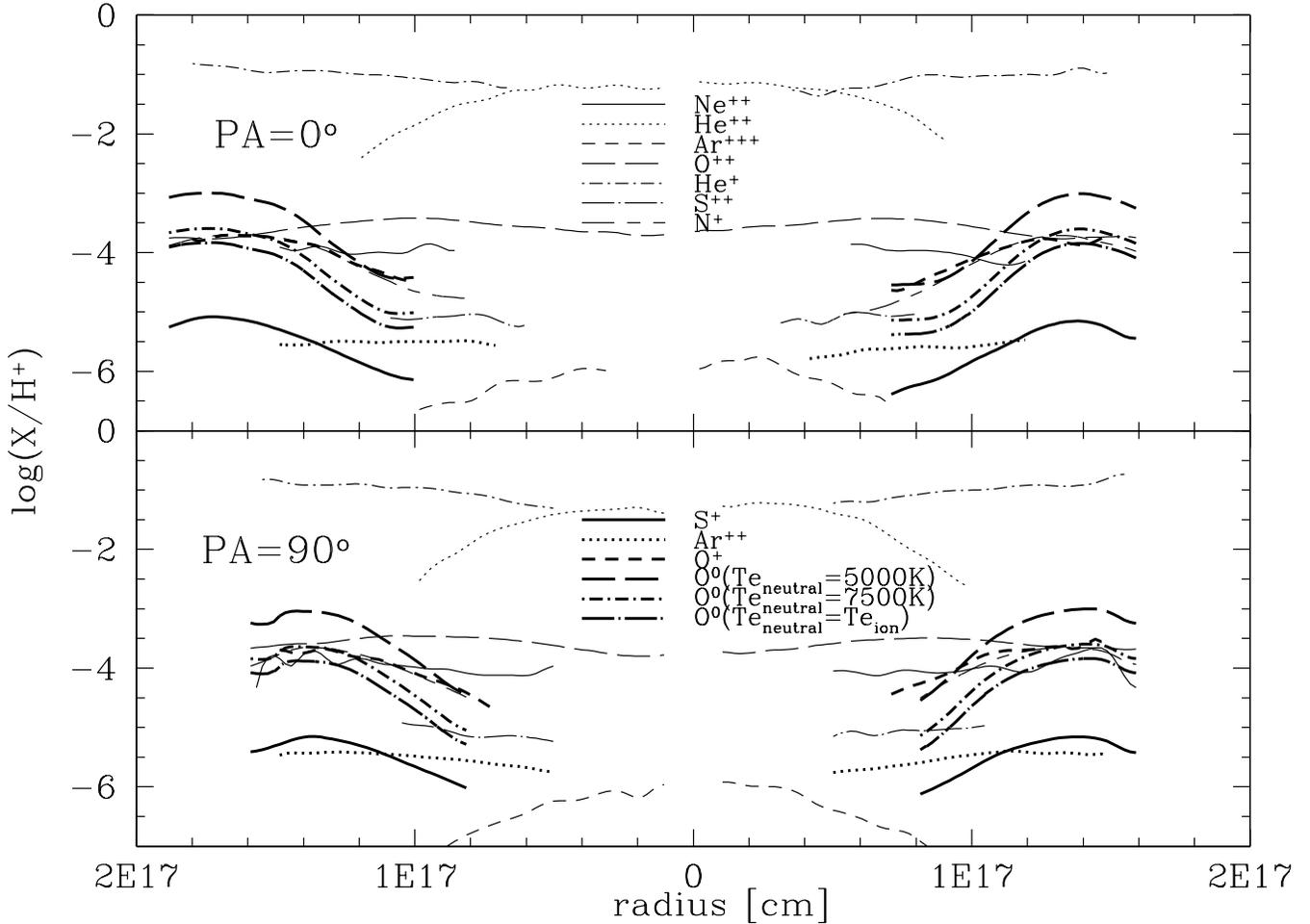} 
\caption{The radial ionization structure of NGC~6565 at PA=0$\degr$ (top) and
PA=90$\degr$ (bottom). Three solutions are shown for O$^0$/H$^+$,
corresponding to Te(neutral)=5000 K, 7500 K and $\simeq$ Te(ionized). The
orientation is as in Fig.~9.}
\end{figure*}

\section{The radial ionization structure}
  
The detailed knowledge of the plasma diagnostics in the (fully ionized +
transition) zone of NGC~6565 allows us to solve the equations of
statistical equilibrium, thus obtaining the point to point ionic
concentrations from the line intensities (Peimbert \& Torres-Peimbert 1971,
Barker 1978, Aller 1984, Osterbrock 1989).

The absolute emission profiles (erg cm$^{-3}$ s$^{-1}$) in the zvpc at 
PA=0$\degr$ and PA=90$\degr$ are shown in Fig.~10.

All the main ionic species are well represented in the echellograms, 
with the exception of O$^+$, whose principal emissions 
($\lambda$3726\AA\/ and $\lambda$3729\AA) fall outside our
spectral range. Thus, we are forced to derive the O$^+$/H$^+$ abundance 
from the $\lambda$7319.92\AA/H$\alpha$ ratio, 
where $\lambda$7319.92\AA\/ is the strongest line of the 
red O$^+$ quartet, whose relative intensities are practically constant in 
the density range here considered (De Robertis et al. 1985). 
Since the ratio $\lambda$7319.92\AA/($\lambda$3726\AA+$\lambda$3729\AA) is in the range 
0.015 to 0.075, mainly depending on Ne (Keenan et al. 1999), the O$^+$/H$^+$ abundance 
is obtained only at, or close to, the line peaks.  
Also, the unfavorable position of $\lambda$3967\AA, at the extreme blue edge
of the frame, makes the quantitative Ne$^{++}$/H$^+$ analysis quite uncertain.

At this point some caveats are in order:
\begin{description}
\item[1)] because of the luminosity drop of the fast evolving central star,
the ionization and thermal structure of NGC~6565 are out of equilibrium: 
with time the H$\alpha$ peak recedes (the ionized nebula zooms
out) and the recombination processes prevail. They are faster in the high ionized species; 
this explains, at least qualitatively, the
reversion of the Zanstra discrepancy reported in Sect. 7.4: T$_{\rm Z}$HI is
larger than T$_{\rm Z}$HeII because H$\beta$ maintains a ``longer memory'' than
$\lambda$4686\AA\/ of the stellar luminosity.

Actually, the post-AGB evolution slows down and the luminosity gradient rapidly
decreases while approaching the white dwarf domain (Sch\"onberner 1979, 1981,
1983; Wood \& Faulkner 1986; Vassiliadis \& Wood 1993, 1994; Bl\"ocker 1995).
Because of the high gas density we can consider NGC~6565 in quasi-
equilibrium: the H$^+$ recombination rate, dH$^+$/dt$\propto$-Ne$^2$, is large 
enough to ensure a short ``relaxation'' of the main nebula (the
recombination time, t$_{\rm rec}$(yr)=1/$\alpha_{\rm B}$Ne$\simeq$1.2$\times$10$^5$/Ne,
is a few dozen years; $\alpha_{\rm B}$=effective recombination coefficient of
hydrogen).

\item[2)] No correction for seeing is applied to the observed emission line
profile; although this causes some inaccuracy for the sharpest radial
distributions (like [OI], [NII] and [SII]), it doesn't modify the general
results here obtained.

\item[3)] No precise information is yet available for the electron temperature
of the patchy, almost neutral gas in the external part of the ``transition''
zone, where the recombination and cooling processes dominate. This mainly
concerns the [OI] $\lambda$6300\AA\/ emission, which is much more sensitive to
the physical conditions than are the [NII] and [OII] lines (Williams 1973).
Thus, we are forced to consider different scenarios.
Three possible solutions are given for O$^0$/H$^+$, 
corresponding to Te(neutral)=7500 K, Te(neutral)=5000 K  and Te(neutral)$\simeq$Te(ionized).
\end{description}

Fig.~11 shows the resulting radial ionization structure of NGC~6565 
at PA=0$\degr$ (top) and PA=90$\degr$ (bottom).
Only two elements (i.e. helium and oxygen) present an
almost complete ionization structure. Note that hydrogen is well
represented in the main nebula, but the lack of the neutral term
appears problematic in the outermost regions, where H$\degr$ abounds (see
Sects.~6.1 and 7.3).

This is confirmed by the (He$^+$/H$^+$ + He$^{++}$/H$^+$) radial
trend: it is approximately constant (0.105$\pm0.005$) up to the
``transition'' zone and later it increases outwards. Such a behavior
is typical of an optically thick PN ionized by a high temperature
central star (the He$^+$ region extending further than the H$^+$ one;
Hummer \& Seaton 1964; Alexander \& Balick 1997). Previous He/H
abundances reported in the literature for NGC~6565 (mean values over
the whole nebula) span in the range 0.095 to 0.108 (Aller et al. 1988;  K\"oppen et al. 
1991; de Freitas Pacheco et al. 1992).

\subsection{Ionization structure for Te(neutral)$\simeq$Te(ionized)}

This corresponds to the common assumption of using a single temperature 
(Te[OIII] and/or Te[NII]) all over the nebula, even in the regions
where recombination starts to be significant.

In this case the value (O$^0$/H$^+$ + O$^+$/H$^+$ +
O$^{++}$/H$^+$) is quite constant across the nebula (but in the
innermost regions, where O$^{+3}$ and even higher ionization stages
dominate). O/H results 6.0($\pm$1.0)$\times$10$^{-4}$, to be compared
with the values of 5.9$\times$10$^{-4}$, 8.7$\times$10$^{-4}$ and
10.9$\times$10$^{-4}$ reported by Aller et al. (1988), K\"oppen et al. (1991) and de 
Freitas Pacheco et al. (1992), respectively.

The most striking feature of Fig. 11 is the low O$^0$/H$^+$ abundance
in the outer regions of NGC~6565. This is a direct consequence of the
``large'' value adopted for Te(neutral), favoring the emissivity of the
forbidden line $\lambda$6300\AA\/ with respect to H$\alpha$.  The
O$^0$/H$^+$ under-abundance indicates a scarce efficiency of the
charge-exchange reaction O$^+$ + H$^0$$\getsto$O$^0$ + H$^+$ (Williams
1973; Aller 1984; Osterbrock 1989), implying that H$^+$$\gg$H$^0$
(in contradiction with the result just obtained from He).

All this is likely due to the unrealistic assumption
Te(neutral)$\simeq$Te(ionized), for the following reasons:

\begin{description}

\item[-] the low value of O$^0$/H$^+$ in the outer regions of NGC~6565
disagrees with a number of observational evidences mentioned before, 
indicating the presence of a large amount of
neutral, dusty and patchy gas around the ionized nebula;

\item[-] H$^+$$\gg$H$^0$ means that the nebula is almost optically thin to 
the UV flux of the star, i.e. the total nebular mass is close to the
ionized nebular mass
(M$_{\rm tot}$$\simeq$M$_{\rm ion}$$\simeq$0.03M$_\odot$). Thus, NGC~6565 should
be an under-massive nebula ejected and excited by a massive central
star, contradicting the theoretical and
observational evidences of a direct relation between M$_{\rm neb}$ and
M$_*$ (Sch\"onberner 1981; Pottasch 1983; Vassiliadis \& Wood 1994;
Boffi \& Stanghellini 1994; Bl\"ocker 1995; Buckley \& Schneider 1995;
Dopita et al. 1996).
\end{description}

\begin{figure}
\centering
\includegraphics[angle=-90,width=9.2cm]{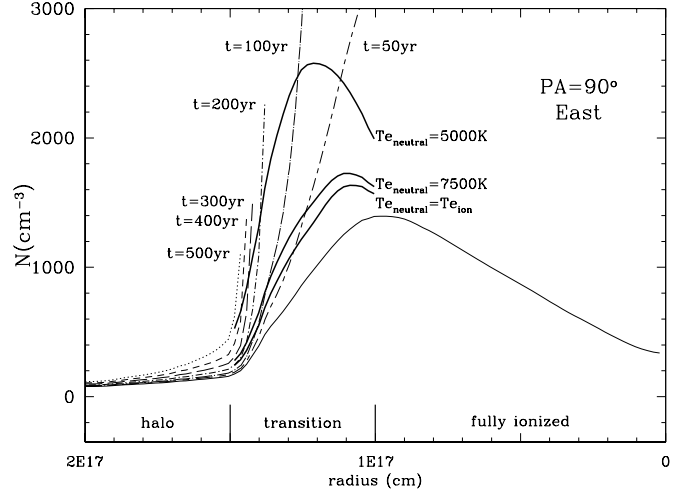} 
\caption{Reconstruction of the (ionized+neutral) radial density profile
in NGC~6565 at PA=90$\degr$ East sector. The orientation
is as in Fig. 9. Thin continuous line: the observed Ne$\simeq$N(H$^+$)
profile (from the H$\alpha$ flux in the zvpc), for $\epsilon_{\rm l}$=1.0. Thick continuous lines: the 
N(H$_{\rm tot}$) distribution in the ``transition'' zone (from Eq. (8)) for three
values of Te(neutral). Dashed and dotted thin lines: the N(H$_{\rm tot}$) profile
in the ``transition zone'' and in the ``halo'' for some representative values
of time elapsed from the start of the recombination (from Eq.~(10)).}
\end{figure}

\subsection{Ionization structure for Te(neutral)$<$Te(ionized)}

Lowering Te(neutral), O$^0$/H$^+$ increases in the outermost part, 
suggesting that charge-exchange reactions dominate
in this H$^0$-rich region (see Fig.~11).

The density profile of H$^+$ has been derived all over the nebula
in Sect.~6.4. The H$^0$ concentration in the ``transition'' zone is 
difficult to estimate. 
An indication can be obtained as follows.

In the general form, the total abundance
(relative to hydrogen) of an element X with atomic number n is given
by:

\begin{equation}
\frac{X}{H} = \frac{\sum_{\rm i=0}^n X^i}{\sum_{\rm j=0}^{1} H^j}
\end{equation} 

For oxygen, neglecting the higher ionization terms, this can be written as:

\begin{equation}
\frac{O}{H} = \frac{O\degr + O^+ +O^{++}}{H\degr + H^+}
\end{equation}

Assuming O/H=constant across the nebula, we obtain for each radial position:

\begin{equation}
\frac{H\degr + H^+}{H^+} \simeq \frac{\frac{O\degr}{H^+} + \frac{O^+}{H^+} + 
\frac{O^{++}}{H^+}}{\frac{O}{H}} 
\end{equation}

which provides the H$^0$ distribution in the 
``transition'' zone of NGC~6565, since O/H is derived from the ``fully 
ionized'' nebula. 

The impasse caused by the uncertainties on O$\degr$/H$^+$, due to Te(neutral), can be overcame by 
linking the ``transition'' zone with the ``halo'', the outermost, 
very low electron density region which is in a 
recombination phase since the ``recent'' luminosity drop of the star. 

The question is: how recent?  A lower limit of about 100 years comes
from the Curtis (1918) description of NGC~6565 (''No central star can
be seen`` with the 0.91m Crossley Reflector).  Moreover, the value
M$_*$$\simeq$0.65 M$_\odot$ adopted for the stellar mass (see
Sect. 7.4) suggests that the luminosity drop indicatively started
1000 years ago.  This must be considered as an upper limit, since the
thin--thick transition did not occur at the beginning of the
stellar drop, but during the decline (the exact moment depending on
the nebular mass; for M$_{\rm neb}$$\simeq$0.15 M$_\odot$ the
thin--thick transition started at log L$_*$/L$_\odot$$\simeq$
2.8, i.e. about 400 years ago).

More precise information comes from the observed density profile in the ``transition zone'' and 
in the ``halo''.
The Ne depletion rate for recombination is given by: 
\begin{equation}
dNe/dt = -\alpha_{\rm B} Ne N(H^+) 
\end{equation}
Assuming Ne$\simeq$N(H$^+$) and Te=10000 K, and neglecting the 
recombination delay due to expansion, we obtain:
\begin{equation}
N(H_{\rm tot}) = Ne(t)/[1 - 8.2\times10^{-6} t Ne(t)] 
\end{equation}
which provides N(H$_{\rm tot}$), the density in the ``transition zone'' 
and in the ``halo'' at the start of the recombination phase, once
is known the present Ne(t), the electron density at time t (in years) 
elapsed from the beginning of the recombination process.

This is summarized in Fig.~12 for PA=90$\degr$ East sector, i.e. 
along the apparent minor axis of NGC~6565. 
It shows:
\begin{description}
\item[-] the whole Ne$\simeq$N(H$^+$) profile in the zvpc 
(from the H$\alpha$ flux, for $\epsilon_{\rm l}$=1.0); 
according to the discussion in Sect.~6.5, we adopt
Ne$_{\rm halo}$=Ne$_{\rm halo}$(Fig. 9) / 2, 
\item[-]the N(H$_{\rm tot}$) distribution in the ``transition'' zone for Te
(neutral)$\simeq$Te(ionized), 5000 K and 7500 K (from Eq. (8)), 
\item[-] the density profile, N(H$_{\rm tot}$), in the ``transition zone'' and 
in the ``halo'' for some representative values of t (from Eq. (10)). 
\end{description}

In spite of the heavy assumptions (in particular: $\alpha_{\rm B}$ is a function
of Te and $\epsilon_{\rm l}$$<$1.0 in the real nebula), Fig. 12 suggests that:
\begin{description}
\item[-] the recombination of the main nebula indicatively started 
about 400 years ago;
\item[-] the nebular mass is at least 0.10 M$_\odot$, hence
most of the gas is presently neutral (M$_{\rm ion}=0.03 \pm 0.01$,
Sect.~7.3);
\item[-] the electron temperature in the external, almost neutral regions 
is lower than in the ``ionized'' nebula (5000 K$<$Te(neutral)$<$ 7500 K).
\end{description}

Note that the chronological sequence defined by the intersection of the 
different curves in the ``transition zone'' underestimates t for t$<$100--200 years, 
since we have neglected the contribution of the photoionization in the innermost regions 
of the ``transition zone''. This mainly concerns the t=50yr curve, which considerably 
shifts to the right.

As already noticed in Sect. 6.1, an unrealistic situation occurs for Te
(neutral)$\simeq$5000 K: the complete H$\alpha$/H$\beta$
distribution observed in the spectral maps is due to Te radial
variations. This implies that c(H$\beta$)$_{\rm local}$$\simeq$0 and H$^+\gg$H
$^0$ (an evident nonsense stressing the urgency of a more sophisticate
photo-ionization model).

The overall (fully ionized+transition+halo) structure of NGC~6565 closely
recalls the recombination phase of Tylenda's (1983, 1986) models: only the
innermost nebula is ionized by the fading UV luminosity of the star,
whereas the outermost part, unattainable by the direct radiation, simply ``
remembers'' the UV flux received just before the luminosity drop. In this
scenario, the three-zones division scheme proposed in Sect.~6.5 marks two
distinct  evolutionary times of the star: the ``fully ionized''  nebula
refers to the present low luminosity phase, the ``halo'' to its past glory.

The ``transition'' zone is a special case; on the whole it is recombining,
but the process has been gradual (the outermost regions are the first to
recombine), and decelerated (the luminosity gradient of the star decreases
in time).
Moreover, following Tylenda (1986) and Marten \& Szczerba (1997)
the high ionization species recombine faster than the low ones, and 
for a given ion, the shorter the wavelength of the transition, the faster 
is the decline of the respective emission.

The electron temperature behavior in the ``transition'' zone reflects the
rate of change in kinetic energy of (ions+atoms+electrons) and is the net
result of two contrasting processes: the recombinations 
favor an increase of Te since they tend to remove slow electrons, whereas the 
losses by forbidden lines lower Te (Aller, 1984). All
this, added to the inhomogeneous structure of the outermost layers, makes
the situation quite complex. Following the time dependent ionization
calculations by Harrington \& Marionni (1976), Tylenda (1986) and 
Marten \& Szczerba (1997), the electron temperature gradually 
decreases in time.
In short, the ``transition'' zone constitutes a spatial and temporal 
link between the outer edge the ``fully ionized'' nebula and the
recombining ``halo''.

\begin{table}
\caption{Total chemical abundances in NGC~6565}
\begin{tabular}{l}
\hline
\\
He/H=0.105($\pm$0.003)\\
O/H=6.0($\pm$1.0)$\times$10$^{-4}$\\
N/H=1.5($\pm$0.3)$\times$10$^{-4}$\\
Ne/H=1.6($\pm$0.4)$\times$10$^{-4}$\\
S/H=6.3($\pm$1.0)$\times$10$^{-6}$\\
Ar/H=1.8($\pm$0.4)$\times$10$^{-6}$\\
\\
\hline
\end{tabular}
\end{table}

The contracting phase of NGC~6565 will last only a few
hundred years: with time, the dilution factor due to the expansion will
overcome the slower and slower stellar decline, thus leading to a gradual
re-growing of the ionization front.

Actually, we cannot completely rule out the possibility
that the re-growing already takes place in NGC~6565. According to the
classical Str\"omgren model of nebular ionization, the 
contraction--expansion equilibrium occurs for $-1/3 d(ln $L$_*)/$dt$=2/$t$_{\rm dyn}$, 
i.e. it depends on the present rate of luminosity decline of the star.
The evolutionary tracks for hydrogen burning post-AGB stars by 
Sch\"onberner (1981), Wood \& Faulkner (1986), Vassiliadis \& Wood (1994) and
Bl\"ocker (1995) give contrasting results.

Although a clear answer needs first and second epoch HST imagery in the main nebular emissions, 
our feeling is that NGC~6565 is still contracting, 
because of the observed reversion of the Zanstra discrepancy (or, at least, it seems to be 
still contracting, due to the time delay in the nebular adjustement
to the changing UV luminosity of the star).

We conclude this Section with the total chemical abundances (Table 2). 
They are derived in the usual way, i.e. multiplying the observed
ionic abundances for the corresponding ICFs, the correcting
factors for the unobserved ionic stages (obtained both empirically, Barker
1983, 1986, and from interpolation of theoretical nebular models, Shields
et al. 1981, Aller \& Czyzak 1983, Aller 1984, Osterbrock 1989). Following
the critical analysis by Alexander \& Balick (1997), we have considered the
total line intensities (i.e. integrated over the whole spatial profile and
the expansion velocity field).

In general, the chemical abundances of Table 2 are in satisfactory
agreement with the values reported in the literature 
(Aller et al. 1988;  K\"oppen et al. 
1991; de Freitas Pacheco et al. 1992), but for
the N overabundance suggested by Aller et al. (1988) which
is not confirmed here.

\section{The photo-ionization model}

\begin{table}
\caption{Input parameters of the model nebula}
\begin{tabular}{ll}
\hline
\\
Radial density profile & N(H$_{\rm tot}$) in top panel of Fig.~13 \\
                       & (cfr. Sect. 8.2) \\
Filling factor         & 1 \\
Chemical abundances:   & \\
~~ C, Si, Cl           & Aller et al. (1988)\\
~~ He, N, O, Ne, S, Ar & this paper\\
~~ other elements      & PN (CLOUDY default)\\
Dust                   & PN (CLOUDY default)\\
Star                   & blackbody with T$_*$=120000 K and \\
                       & logL$_*$/L$_\odot$= 2.0\\
\\
\hline
\end{tabular}
\end{table}
 
\begin{figure*}
\centering
\includegraphics[width=\textwidth]{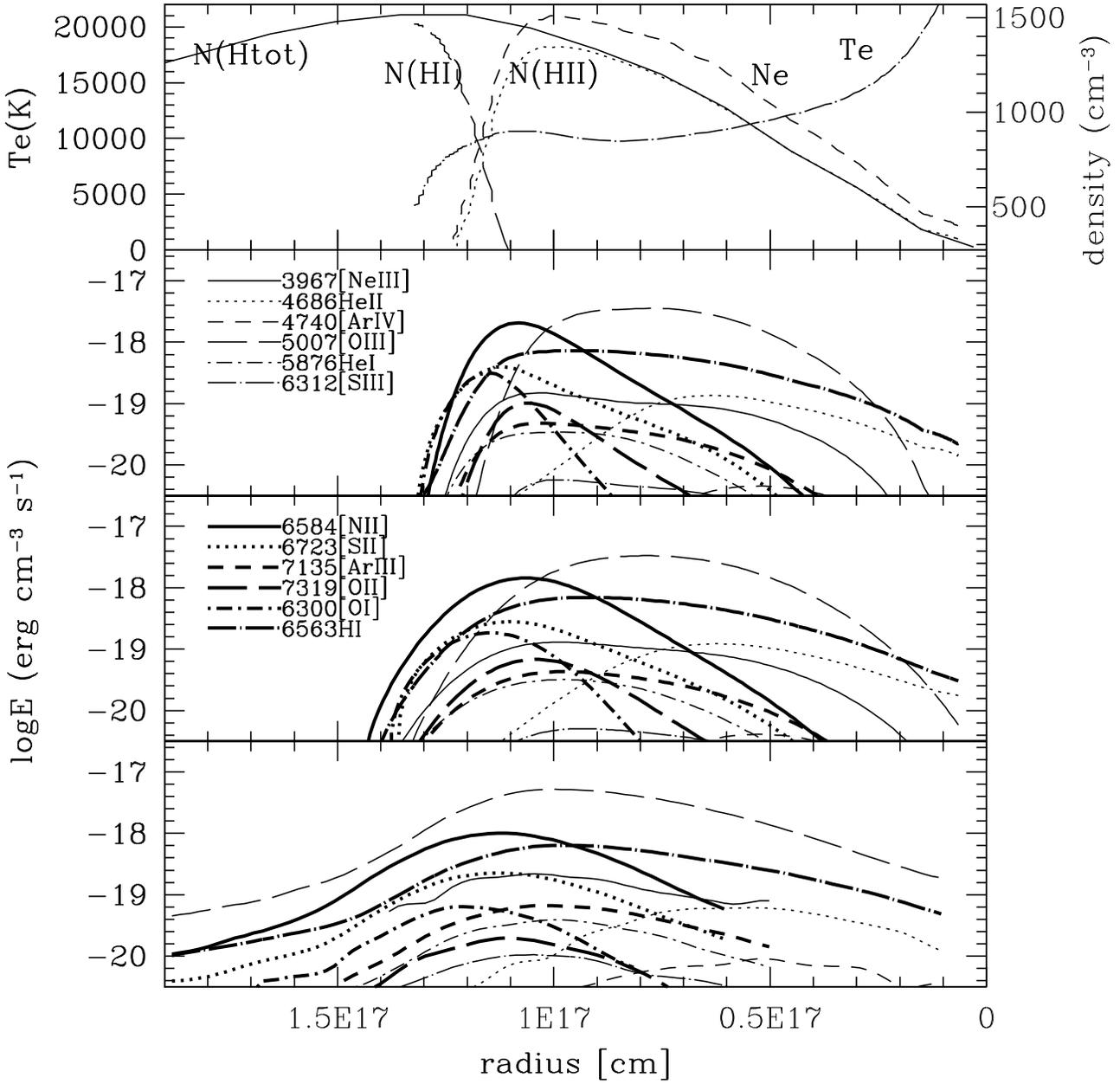} 
\caption{Model nebula vs. NGC~6565 at PA=90$\degr$, East sector. Top panel:
radial profiles of Te (left ordinate scale) and Ne, N(H$^+$), N(H$\degr$) 
and of the input density N(H$_{\rm tot}$)  (right ordinate scale) for the model nebula in the case 
$\epsilon_{\rm l}$=1.0.
Second panel: absolute radial flux distribution (erg s$^{-1}$ cm$^{-3}$) of
the model nebula in the main emissions. Third panel: same as in the second
panel, after convolution for a seeing of 0.65 arcsec, reproducing the
observational conditions of the true nebula. Bottom panel: absolute flux
distribution in the main emissions of NGC~6565 at PA=90$\degr$, East sector
(cfr. Fig. 10).}
\end{figure*}  

In order to test the results obtained for the physical conditions and the
ionization structure, and to disentangle the controversial points of the
foregoing Sections, we have applied the photo-ionization code CLOUDY 
(Ferland et al. 1998) to a mass of gas having the same density distribution
and chemical composition of our PN, and a powering source similar to
the central star of NGC~6565.

The usual caveat: CLOUDY is a ``steady-state'' model. 
It works in the ``fully ionized'' part of NGC~6565,
which is in quasi-equilibrium (see Sect.~8), but it cannot depict the
external, recombining regions originated by the luminosity drop of the
central star.

We consider only a single sector of a single PA 
(East sector of PA=90$\degr$, i.e. 
along the apparent minor axis) as representative of the whole equatorial 
regions of NGC~6565 identified by the zvpc.

The input parameters of the model nebula are given in Table 3. The adopted
N(H$_{\rm tot}$) profile corresponds to the ``fully ionized'' part of N(H$^+$) 
at PA=90$\degr$ East sector, for $\epsilon_{\rm l}$=1 (see Fig.~9). It is
arbitrarily extended to the ``transition'' and the ``halo'' in order to
obtain a nebular mass of about 0.30 M$_\odot$. The choice of different N(H)
profiles in the outermost regions does not modify the resulting ionization
structure, since these layers cannot be reached by the UV stellar flux.
Concerning the ionizing source, we use a blackbody distribution having the
Zanstra HeII temperature and luminosity of the central star of NGC~6565 
(see Sect.~7.4). 
Once it is convolved for a seeing of 0.65 arcsec, the line emissivity (per unit filling factor) 
of the model nebula at 2.0 Kpc can be compared with the emission per unit volume 
presented in Fig.~10.
The results are shown in Fig.~13.

The steady-state model nebula convolved for seeing (third
panel in Fig.~13) closely reproduces the ``fully ionized'' part of NGC~6565
(bottom panel) up to R$\simeq$1.0$\times$10$^{17}$ cm. An even better fit might
be obtained through a modest adjustment of the chemical abundances. 
The discrepancies begin in the ``transition'' zone and increase outward (``halo'').

The comparison of the model vs. observed nebula offers an encouraging
support to the results obtained in the previous Sections. In particular, Fig. 13 confirms 
the peculiar properties of [ArIV] at $\lambda$4740\AA, as pointed out in 
Sect. 4: the structure of Ar$^{+3}$ (I.P.=40.7 eV) mimics that of higher 
ionization species since it is present in the 
interval (40.7--59.8) eV, the upper limit being the ionization potential of Ar$^{+4}$, 
i.e. it overlaps the He$^{++}$ zone.  

Neon shows the opposite behaviour of Argon: Ne$^{++}$  is expected to within the (40.9--63.4) 
eV range, whereas in Fig. 13 the [NeIII] emission at $\lambda$3967\AA\/ extends to the 
medium-low ionization region (see also Sect. 4 and Figs. 5 and 10). We have verified 
on a number of model nebulae: the abnormal extension of Ne$^{++}$ only occurs in the 
presence of a hot central star. According to Williams (1973), Butler et al. 
(1980), Shields et al. (1983) and 
Clegg et al. (1986), a large number of stellar photons 
with energies well above the HI ionization threshold maintains a high 
degree of ionization in the nebula right up to the point where H$^o$ prevails, and the 
charge-exchange reaction Ne$^{+3}$+H$^o\getsto$Ne$^{++}$+H$^+$ populates the $^5$P, $^3$D$^o$ 
and $^3$P levels of Ne$^{++}$. The ground state is then reached via radiative decay, and 
the collisionally excited lines at $\lambda$3868\AA\/ and  $\lambda$3967\AA\/ are emitted. 

Concerning the electron temperature, the radial behavior of the model
nebula (Fig.~13, top panel) closely reproduces the observed one (Fig.~9,
middle panel) up to the ``transition'' zone. The external, fast Te drop of
the model nebula clearly refers to a steady-state situation, which does not
take into account the presence of the broad recombining region in NGC~6565.

Notice the quantitative agreement between E(H$\alpha$) of the model nebula,
convolved for seeing, and the observed one. This is a consequence
of the clumpiness found in Sect.~7.2 (i.e.
$\epsilon_{\rm l}$$\simeq$0.25).
In fact, the former is obtained from the N(H$_{\rm tot}$) profile 
under the assumption $\epsilon_{\rm l}$=1, 
whereas in the real nebula 
N(H$^+)=\epsilon_{\rm l}^{1/2} \times $Ne[SII]$=0.5 \times $Ne[SII].

In Fig.~13 the most striking discrepancy concerns the radial 
O$^{++}$ profile: in NGC~6565 (bottom panel) it closely reproduces 
the H$^+$ one, as expected of an optically thick nebula powered by a high 
temperature central star (Flower 1969, Williams 1973), while
in both the model nebula and the convolved model nebula (second and third panel, respectively)
H$^+$ extends further than O$^{++}$ and the outermost ionized regions are O$^{++}$-depleted.

We stress that the application of the photo-ionization code CLOUDY 
to the other PA of NGC~6565 gives results 
coinciding with the ones just presented for PA=90$\degr$, East sector.

We conclude that the nebular modeling is a powerful 
instrument for interpreting the observed 
properties of each PN. In the specific case of NGC~6565, 
the nebula is so peculiar that it cannot be completely 
reproduced even by a detailed and 
sophisticate code like CLOUDY.
An evolving model is required, which takes into account the nebular 
reaction to the rapidly changing UV luminosity of the star.

NGC~6565 is not a unique case. The recombination phase must be quite common
in PNe, almost ineluctable in the presence of a massive, i.e. fast
evolving, central star. Tylenda (1986) lists a dozen well-studied objects,
including the famous NGC~7027, NGC~7293, NGC~6853 and NGC~6720, and Corradi
et al. (2000) suggest that a recombining halo is present in NGC~2438, but
many other candidates are contained in the Acker et al. (1992) and Kohoutek
(2000) catalogues on the basis of their optical appearance, spectroscopic
characteristics and star faintness. 
Notice that the combination of large
optical extension and high surface brightness makes the Ring nebula the
ideal laboratory for analyzing in detail the recombination process in PNe.

\begin{figure*}
\caption{Opaque reconstruction of NGC~6565 in H$\alpha$ at Ne$\simeq$300 cm$^{-3}$ 
(for $\epsilon_{\rm l}$=1, upper part of each panel) and Ne$\simeq1000$ cm$^{-3}$ (bottom part), 
as seen from 13 directions separated by 15$\degr$. The line of view
is given by ($\theta,\psi$), where $\theta$ is the zenith angle and $\psi$ the
azimuthal angle; the upper-right image is the rebuilt-nebula seen from the
Earth, i.e. from (0,0). Each horizontal couple represents a ``direct''
stereoscopic pair, allowing the reader to have 12 3-D views of the nebula in as
many directions, all together covering a straight angle. We repeat here the
instructions given by Ragazzoni et al. (2001): to obtain the three-dimensional
vision, look at a distant object and slowly insert the figure in the field of
view, always maintaining your eyes parallel. Alternatively, you can use the two
small dots in the upper part of the figure as follows: approach the page till
the two dots merge (they appear out of focus); then recede very slowly, always
maintaining the two dots superimposed, till the image appears in focus.}
\end{figure*}   

\section{The spatial structure}

\subsection{Tomography}
The reconstruction of the emission line structure in the nebular
slices covered by the slit was introduced by Sabbadin et al. (2000a, b) and
Ragazzoni et al. (2001). In the case of NGC~6565 we have selected $\lambda$4686\AA\/ 
of HeII, $\lambda$5007\AA\/ of [OIII] and $\lambda$6584\AA\/ of [NII]
as representative of the high, mean and low ionization regions,
respectively, and H$\alpha$ as a marker of the whole ionized gas
distribution.

The spectral images of the forbidden lines were deconvolved for seeing, 
spectral resolution and thermal motions, while also fine structure was considered 
for the recombination lines. They were de-projected using the
expansion law derived in subsect. 7.2, i.e. adopting 
Vexp/r=5.6 km s$^{-1}$ arcsec
$^{-1}$. The resulting tomographic maps (quite similar to the spectral
images shown in Figs. 4 and 7) were assembled with the procedure described
by Ragazzoni et al. (2001), thus obtaining the spatial structure presented
in the next Section.

\subsection{3-D morpho--kinematical structure}

\begin{figure*}
\centering
\caption{Same as Fig. 14, but for [NII], showing the faintest and the brightest (upper and lower 
part of each panel, respectively) low ionization regions of NGC~6565.
}
\end{figure*}  

In order to render the 3-D structure, we adopt the method introduced by
Ragazzoni et al. (2001): a series of images showing the nebula from different
directions, separated by 15$\degr$. Each couple of images constitutes a
stereoscopic pair, illuding the reader to see NGC~6565 in 3-D.

\begin{figure*}
\caption{[OIII] and [NII] appearance of NGC~6565 from 6 directions separated 
by 30$\degr$, showing the different morphologies when changing the line of view.  The upper-right panel 
($\theta,\psi$=0,0) corresponds to the rebuilt-nebula seen 
from the Earth (to be compared with Fig. 1). The scale is the same of Figs. 14 and 
15. Remember that projection($\theta$,$\psi$)=projection($\theta\pm$180$\degr$,$\psi\pm$180$\degr$).}
\end{figure*}  

The opaque reconstruction in H$\alpha$ is shown in Fig. 14 for two absolute flux cuts: 
log E(H$\alpha$)=-19.55 erg s$^{-1}$ cm$^{-3}$ (upper part of each panel) and 
-18.50 erg s$^{-1}$ cm$^{-3}$ (lower part). Since E(H$\alpha$)=h $\nu$ $\alpha^{eff}_{\rm H\alpha}$
Ne N(H$^+$) $\epsilon_{\rm l}$, the adopted cuts correspond to Ne$\simeq$300 cm$^{-3}$ (upper) and Ne$\simeq$1000 
cm$^{-3}$ (lower) (for Te=10$^4$ K and $\epsilon_{\rm l}$=1).
   
The complexity of the nebular structure in the faint, low ionization polar cups is more 
evident in Fig. 15, referring to 
$\lambda$6584\AA\/ of [NII] at the two cuts: log E($\lambda$6584\AA)= -19.40 erg s$^{-1}$ cm$^{-3}$ (upper 
part of each panel) and -18.40 erg s$^{-1}$ cm$^{-3}$ (lower part).
In this case E($\lambda$6584\AA)$\propto$Ne N(N$^+$), thus a precise Ne cut cannot be assigned: 
Fig. 15 simply represents the ``faintest'' and the ``brightest'' low ionization regions, respectively.

NGC~6565 in the high ionization HeII emission at $\lambda$4686\AA\/ (not shown
here for reasons of space) consists of a compact, almost tubular structure recalling the H$\alpha$
one (high cut) presented in Fig. 14. Moreover, the spatial form in [OIII] at
$\lambda$5007\AA\/ closely reproduces the H$\alpha$ one (Fig. 14), as expected of
the similarity of the spectral images and the radial profiles (see Sects. 2, 4
and 8).
Finally, the distribution of the neutral gas in the transition zone of NGC~6565
(obtained from Eq.~(8) using the approximation 
H$_{\rm tot}$/H$^+$$\simeq$O$\degr$/O$^{++}$ and assuming Te(neutral gas)=6000 K 
and Te([OIII])=11000 K) mimics the
N$^+$ structure shown in Fig. 15, although  peaked a bit further away.

In short, NGC~6565 is an inhomogeneous triaxial ellipsoid seen almost pole-on (major axis=10.1 
arcsec projected in (5,7), intermediate axis=7.1 arcsec in ($-85$,7), minor 
axis=6.0 arcsec in (5,$-83$)). The matter along and close to the major axis was swept-up by some 
accelerating agent (high velocity wind? ionization? magnetic fields?), forming two fast, 
asymmetric and faint polar cups projected in (18,7; i.e. they are lightly misaligned with respect 
to the major axis) and extending up to about 10.0 arcsec from the star. The optical 
nebula is embedded in a large cocoon of almost neutral gas, which 
is the result of the recombining processes occurred during the luminosity decline of the powering 
star.

The [OIII] and [NII] appearance of NGC~6565 seen from different directions is
illustrated in Figs. 16. It gives a representative sample of the possible
morphologies assumed by our nebula when changing the line of view, and confirms
its structural similarity with NGC~6886 (suggested in Sect. 3), as well with 
K~3-92, M~1-7, M~1-59, M~1-66, M~2-40 and M~2-53 (see the imagery catalogues by
Schwarz et al. 1992, Manchado et al. 1996, and Gorny et al. 1999).

Also the observed kinematics depends on the line of view, as shown in Table 4 
(upper panel). NGC~6565 can be either a compact, roundish, fast expanding object
or a bipolar, elongated, slow one with all the intermediate cases. The
morpho--kinematical range further spreads out in the case of a ``butterfly'' PN,
like Hb~5, Mz~3 and NGC~6537.

Decreasing of the spatial resolution exacerbates the problem. For example, by
putting NGC~6565 at the distance of the Large Magellanic Cloud (i.e. 50 Kpc; Kovacs 2000 
and references therein) we obtain the results of Table 4, bottom
panel.
In this case the echellograms completely lose the spatial information due to
the nebula compactness. 2Vexp[OIII](peak separation) cannot be measured any
longer (in Table 4 it has been replaced by 2Vexp[OIII](FWHM)) and the
morphological classification is quite uncertain.

\begin{table*}
\caption{Morpho-kinematical characteristics}
\begin{tabular}{lllllll}
\hline
\\
\multicolumn{7}{c}{NGC 6565 at 2.0 Kpc}\\
\\
Projection      &  2Vexp[OIII]    &  2Vexp[NII]    &  2Vexp[OIII]           &2Vexp[NII]            & morphology & 
apparent size \\
      & \multicolumn{2}{c}{(peak separation)} & \multicolumn{2}{c}{(10\% maximum intensity)} &            &     \\
      & [km s$^{-1}$]     & [km s$^{-1}$]    &  [km s$^{-1}$]           & [km s$^{-1}$] &            &     [arcsec]\\
\\
along the major axis    &  57($\pm$2) & 61($\pm$2) & 90($\pm$4) & 105($\pm$4) & elliptical & 10.0$\times$8.5\\
along the intermediate axis &42($\pm$2)& 46($\pm$2)& 50($\pm$4) & 77($\pm$4)  & bipolar    & 20.0$\times$8.5\\
along the minor axis    & 34($\pm$2)   & 39($\pm$2)& 40($\pm$4) & 62($\pm$4)  & bipolar    &20.0$\times$10.0\\
\\
\\
\multicolumn{7}{c}{NGC 6565 at the distance of the Large Magellanic Cloud}\\
\\
Projection      &  2Vexp[OIII]    &  2Vexp[NII]    &  2Vexp[OIII]           &2Vexp[NII]            & morphology & 
apparent size \\
      & (FWHM) & (peak separation) & \multicolumn{2}{c}{(10\% maximum intensity)} &            &     \\
      & [km s$^{-1}$]     & [km s$^{-1}$]    &  [km s$^{-1}$]           & [km s$^{-1}$] &            &     [arcsec]\\
\\
along the major axis   & 57($\pm$4)    & 47($\pm$4)& 80($\pm$4) &86($\pm$4)   &    -        &0.40$\times$0.34\\
along the intermediate axis &44($\pm$4)&36($\pm$4)& 60($\pm$4) & 66($\pm$4)  &     -        & 0.80$\times$0.34\\
along the minor axis  &36($\pm$4) & 31($\pm$4) & 48($\pm$4) & 55($\pm$4) &         -        &0.80$\times$0.40\\
\\
\hline
\end{tabular}
\end{table*}

All this evidenziates the difficulties implied in the analysis of the high dispersion
spectra and questions the reliability of the interpretations, evolutionary correlations and
spatio-kinematical models based on echellograms at inadequate spatial and spectral
resolutions.

\section{Discussion and Conclusions} 

The long travel dedicated to the study of NGC~6565 is over.
We started with the kinematical properties, passed through the physical conditions and 
arrived to the distance and radius of the nebula and to the temperature and luminosity 
of the star.
The radial ionization structure was later explored, using both the ``classical'' method and 
the photo-ionization code CLOUDY, and the spatial structure investigated by 
assembling the tomographic maps.  

In nuce: NGC~6565 is a young (2000--2500 yr), patchy, optically thick
ellipsoid with extended polar cups, seen almost pole-on. It is in a
recombination phase, because of the luminosity drop of the massive
powering star, which is reaching the white dwarf domain. The stellar
decline started about 1000 years ago, but the nebula remained optically
thin for other 600 years before the recombination phase occurred 
(thus, at Galileo Galilei times NGC~6565 was at high excitation, 
a bit larger and much brighter than at present). In the near future the 
ionization front will re-grow, since the dilution factor due to the 
expansion will prevail on the slower and slower luminosity decline.

The nebula, at a distance of 2.0 Kpc, can be divided into three radial zones:

- the ``fully ionized'', extending up to 0.029--0.035 pc at the equator ($\simeq$0.050 pc at the poles), 

- the ``transition'', up to 0.048--0.054 pc ($\simeq$0.080 pc at the poles),

- the recombining ``halo'', detectable up to 0.110 pc.
  
The ionized mass ($\simeq$0.03 M$_\odot$) is only a fraction of the total mass 
(M$_{\rm tot}$$\ge$0.15 M$_\odot$), 
which has been ejected by an equatorial enhanced superwind of 4($\pm$2)$\times$10$^{-5}$ 
M$_\odot$ yr$^{-1}$ lasted for 4($\pm$2)$\times$10$^3$ years. 

From all the points of view NGC~6565 results to be an exciting laboratory 
which deserves further attention.
Amongst the many and important aspects left unresolved by this paper we 
mention:

\begin{description}
\item[-] the neutral gas distribution and kinematics. Detailed radio and infrared observations are 
highly advisable; for instance, the spectroscopically resolved structure in the H$_2$ emission at 
2.122 $\mu$m, which is a signpost of the bipolar structure (Kastner et al. 1996); 
\item[-] the filling factor. We obtain $\epsilon_{\rm l}$=0.25 in the external, low ionization 
regions by comparing Ne(H$\alpha$) and Ne[SII]. The same analysis should be performed in the 
internal, high ionization parts using  [ArIV] ($\lambda$4711\AA/$\lambda$4740\AA) and [ClIII] 
($\lambda$5517\AA/$\lambda$5537\AA). Deeper echellograms are required;  
\item[-]  the faint and fast (shocked?) polar cups, overlooked in our study mainly focused on the 
equatorial regions; 
\item[-] the mechanisms and physical processes forming and shaping a PN like NGC 6565.
This point appears problematic, mainly because of the huge number of
models proposed for bipolar PNe.

They include: single AGB progenitors (+ fast wind + magnetic fields + photoionization), 
a planetary system,
a binary system undergoing common envelope evolution, a close companion presenting a 
dense accretion
disk (for details, see Frank 1999 and Soker \& Rappaport 2001).

In all these models the nebular shaping is a slow and gradual process occurring 
in a time scale comparable to 
the PN life, whereas the superior spatial resolution of the HST imagery has recently revealed that 
bipolarity is a common feature already in proto-PNe (see Sahai \& Trauger 1998, and references therein). 
This, on the one hand means that some imprint agent acts in the late AGB and/or early 
post-AGB phase, driving most of the subsequent nebular evolution, on the other hand stresses the 
inadequacy of the current models.
\end{description}

In addition to the specific analysis of our nebula, the aim of the present work was
the search for a satisfying reduction procedure allowing us to exploit the huge amount of  
information contained in the high resolution spectra. They essentially cover two 
wide fields: kinematics (including expansion velocity, tomography and spatial structure), and physical 
conditions (radial profile of Ne, Te, ionization, chemical abundances etc.).
The difficulties connected to the determination of the different parameters are perfectly synthesized in the 
Aller's (1994) sentence: ''A nebula is a three-dimensional structure for which we obtain a 
two-dimensional projection''. 
With the introduction of the 3-D methodology, the second half of the sentence becomes: ``... for which 
we obtain the three-dimensional structure''. 

The accuracy of the 3-D reconstruction is defined by the ``relative'' spectral and 
spatial resolutions (Ragazzoni et al. 2001). The first is given by RR=$\Delta$V/Vexp, $\Delta$V being 
the spectral resolution. RR mainly depends on the intrumentation, since the expansion velocity range of 
the PNe is quite sharp (indicatively: 20 km s$^{-1}$$<$Vexp$<$40 km s$^{-1}$). 
The relative spatial resolution, SS=d/$\Delta$d 
(d=angular extent, $\Delta$d=seeing+guiding), essentially depends on the target, due to the 
large spread of apparent sizes exhibited by the PNe.
In the case of a compact object like NGC~6565, the ESO NTT+EMMI echellograms give RR$\simeq$5 and 
SS$\simeq$10; a significant increase of the spatial resolution is expected for the largest nebulae 
of the sample (like NGC~7009 and NGC~6818).

In conclusion, we believe that the crucial point for the physical interpretation of the PNe is the 
detailed knowledge of the radial density profile, now obtainable from the H$\alpha$ flux distribution in 
the zero-velocity pixel column. It opens the possibility of creating a realistic model 
for each expanding nebula (this includes: PNe, nova and supernova remnants, shells around Population I 
Wolf-Rayet stars, nebulae ejected by symbiotic stars, bubbles surrounding early spectral type main 
sequence stars etc.).    

\begin{acknowledgements}
We greatly appreciated the suggestions, criticisms and encouragements by Romuald Tylenda, Arsen Hajian, Gary 
Ferland, Detlef Sch\"onberner and Vincent Icke.
\end{acknowledgements}

\end{document}